\documentclass[aps,prd,twocolumn,superscriptaddress,floatfix,showpacs,showkeys,preprintnumbers]{revtex4}

\usepackage[utf8]{inputenc}
\usepackage[T1]{fontenc}
\usepackage{slashed}
\usepackage{times}

\usepackage{amssymb,amsfonts,amsmath,amsthm}
\usepackage{dsfont,bbm}
\usepackage{dcolumn}

\usepackage{epsf}
\usepackage{epsfig}
\usepackage{graphicx}
\usepackage[caption=false]{subfig}
\usepackage{epstopdf}
\usepackage{dsfont}

\usepackage{multirow}

\usepackage{slashed}

\usepackage[active]{srcltx}

\usepackage[usenames]{color}

\usepackage{wick}

\newcommand{\bit}[1]{\boldsymbol{\mathcal{#1}}}

\newcommand{\der}{\stackrel{\leftrightarrow}{D}}
\newcommand{\derleft}{\stackrel{\leftarrow}{D}}
\newcommand{\derright}{\stackrel{\rightarrow}{D}}

\newcommand{\sderleft}{\not\stackrel{\leftarrow}{D}}
\newcommand{\sderright}{\not\stackrel{\rightarrow}{D}}

\begin{document}

%%%%%%%%%%%%%%%%%%%%%%%%%%%%%%%%%%%%%%%%%%%%%%%%%%%%%%%%%%%%%%%%%%%%%%%%%%%%
\title{Transition form factors $\gamma^*\gamma\to\eta$ and $\gamma^*\gamma\to\eta'$ in QCD}

%\begin{titlepage}

\date{\today}

\author{S.S.~Agaev}
\affiliation{Institut f\"ur Theoretische Physik, Universit\"at
   Regensburg, D-93040 Regensburg, Germany}
\affiliation{Institute for Physical Problems, Baku State University,
 Az--1148 Baku, Azerbaijan}
\author{V.M.~Braun}
\affiliation{Institut f\"ur Theoretische Physik, Universit\"at
   Regensburg, D-93040 Regensburg, Germany}
\author{N.~Offen}
\affiliation{Institut f\"ur Theoretische Physik, Universit\"at
   Regensburg, D-93040 Regensburg, Germany}
\author{F.A.~Porkert}
\affiliation{Institut f\"ur Theoretische Physik, Universit\"at
   Regensburg, D-93040 Regensburg, Germany}
\author{A.~Sch{\"a}fer}
\affiliation{Institut f\"ur Theoretische Physik, Universit\"at
   Regensburg, D-93040 Regensburg, Germany}

\begin{abstract}
We update the theoretical framework for the QCD calculation of
transition form factors $\gamma^*\gamma\to\eta$ and $\gamma^*\gamma\to\eta'$ at 
large photon virtualities including full next-to-leading order analysis of 
perturbative corrections, the charm quark contribution, and taking into account
$SU(3)$-flavor breaking effects and the axial anomaly contributions to the 
power-suppressed twist-four distribution amplitudes. 
The numerical analysis of the existing experimental data is performed with these improvements. 
\end{abstract}

%\preprint{DESY 08-039, Edinburgh 2008/14, LTH 787}


\maketitle

%\newpage

%%%%%%%%%%%%%%%%%%%%%%%%%%%%%%%%%%%%%%%%%%%%%%%%%%%%%%%%%%%%%%%%%%%%%%%%%%%%%%%%%%%%%%%%%%%%%%%%%%%
%
\section{Introduction}
\label{sec:intro}
%
%%%%%%%%%%%%%%%%%%%%%%%%%%%%%%%%%%%%%%%%%%%%%%%%%%%%%%%%%%%%%%%%%%%%%%%%%%%%%%%%%%%%%%%%%%%%%%%%%%%

During last years features of the light pseudoscalar $\eta$ and $\eta^{\prime}$ mesons, 
their quark-gluon structure and hard processes involving these particles, e.g. electromagnetic transition form
factors (FFs) 
and weak decays $B\to\eta (\eta^{\prime})$, were the subject of numerous experimental and theoretical studies. 
Especially the recent measurements of the electromagnetic transition FFs  $\gamma^*\gamma\to\eta$ and 
$\gamma^*\gamma\to\eta'$ at space-like momentum transfers in the interval $4-40\,\,\mathrm{GeV}^{2}$~\cite{BABAR:2011ad}
and at the very large time-like momentum transfer $112\,\,\mathrm{GeV}^{2}$~\cite{Aubert:2006cy} by the BaBar 
collaboration caused much excitement. These measurements and their comparison to the space-like data 
for $\gamma^*\gamma\to\pi^0$ FF in the similar 
range by BaBar and Belle collaborations~\cite{Aubert:2009mc,Uehara:2012ag} stimulated a flurry of theoretical activity, 
e.g.~\cite{Wu:2011gf,Kroll:2013iwa,Klopot:2013oua,Escribano:2013kba}. 
This debate focuses on the question whether hard exclusive hadronic reactions are under theoretical control, which is highly relevant 
for all future high-intensity, medium energy experiments like, e.g., BelleII and PANDA.  

In the exact flavor $SU(3)$ limit the $\eta$ meson is part of the flavor-octet whereas $\eta'$ is a pure flavor-singlet which
properties are intimately related to the celebrated axial anomaly~\cite{Witten:1978bc,Veneziano:1979ec}. However, it is known
empirically that the $SU(3)$ breaking effects are large and have a nontrivial structure. These effects are usually described in 
terms of a certain mixing scheme that considers the physical $\eta,\eta'$ mesons as a superposition of fundamental (e.g. flavor-singlet and
octet) fields in the low-energy effective theory, see e.g.~\cite{Feldmann:1998vh} and references therein. 
It is not obvious whether and to which extent the approach based on state mixing is adequate for the description of hard processes that are
dominated by meson wave functions at small transverse separations, dubbed distribution amplitudes (DAs), 
however, it can be taken as a working hypothesis to avoid proliferation of parameters.
%the first approximation. 

One particularly important issue is that eta mesons, in difference to the pion, can contain a significant admixture of 
the two-gluon state at low scales, alias a comparably large two-gluon DA.      
Several different reactions were considered in an effort to extract or at least constrain these contributions. 
Non-leptonic exclusive isosinglet decays \cite{Blechman:2004vc} and central exclusive production 
\cite{Harland-Lang:2013ncy} act as prominent probes 
for the gluonic Fock-state since the gluon production diagram enters already at leading 
order (LO). Exclusive semi-leptonic decays of heavy mesons were calculated in the framework of 
light-cone sum rules (LCSRs)~\cite{Ball:2007hb,Offen:2013nma} and $k_T$-factorization \cite{Charng:2006zj}.
From a calculational point of view these decays are simpler but the interesting gluon contribution enters 
only at next-to-leading order (NLO). Numerically it was shown that 
the gluonic contributions to $\eta$ production are negligible while they can reach a few percent in 
the $\eta^\prime$-channel. Up to now experimental data are not conclusive in all these decays, with a vanishing 
gluonic DA being possible at a low scale. On the other hand, a large gluon contribution was advocated in~\cite{Liu:2012ib}
from the analysis of $B_{d}\to J/\Psi \eta^{(')}$ transitions (see also~\cite{Hsu:2007qc}). 

In this paper we consider electromagnetic transition form factors  $\gamma^*\gamma\to \eta, \eta'$ that are the 
simplest relevant processes and are best understood from the theory side. 
Also in this case we will find that the present experimental
data are insufficient to draw definite conclusions. However, the 
forthcoming upgrade of the Belle experiment and the KEKB accelerator~\cite{Adachi:2014kka} that aims to increase 
the experimental data set by the factor of 50,  will allow one to measure transition form factors and related observables 
with unprecedented precision.  

The special role of the transition FFs as the ``gold plated'' observables for the study of meson DAs
is widely recognized. To leading power accuracy in the photon virtuality these FFs can be calculated 
rigorously in QCD in the framework of collinear factorization (pQCD)~\cite{Chernyak:1977as,Radyushkin:1977gp,Lepage:1979zb,Efremov:1980mb}.
The main advantage of transition FFs in comparison to other hard reactions with the same property is that 
the leading hard contribution starts already at tree-level and is not suppressed by the usual perturbative penalty factor  
$\alpha_s/\pi \sim 1/10$. For the leading-twist collinear factorization to hold,  
the pQCD contribution has to win against the power-suppressed (end-point or higher-twist) corrections, and this is expected to happen 
for transition FFs already at moderate photon virtualities that are accessible in present experiments. 
One more advantage is that soft contributions are simpler and can be modelled to a reasonable accuracy using, e.g., LCSRs.

The theory of $\gamma^*\gamma\to\eta^{(')}$ decays is, on the one hand, similar to the QCD description of the 
$\gamma^*\gamma\to \pi^0$ transition FF, but, on the other hand, contains specific new issues due to the two-gluon state
admixture, contributions of heavy quarks, and also potentially large meson mass corrections.   
Our goal is to present a state-of-the-art treatment of these special issues
using a combination of perturbative QCD for the calculation of the leading terms and LCSRs for the estimate of power corrections,
complementing our study~\cite{Agaev:2010aq,Agaev:2012tm} of $\gamma^*\gamma\to\pi^0$.
For earlier work related to this program, see \cite{Agaev:2001rn,Kroll:2002nt,Agaev:2002ek,Agaev:2010zz,Kroll:2013iwa}.

An alternative approach to the calculation of transition form factors makes use of transverse momentum dependent (TMD) meson wave functions 
(TMD- or $k_T$-factorization~\cite{Li:1992nu}). This is a viable technique that has been advanced recently to  
NLO, see e.g. \cite{Hu:2012cp,Li:2013xna} for the electromagnetic pion form factor and $\gamma^*\gamma\to\pi^0$, 
and which can be applied to the $\gamma^*\gamma\to\eta^{(')}$ transitions as well. 
Because of a more complicated nonperturbative input, interpretation of the 
corresponding results in terms of DAs is, however, not straightforward so that we prefer to stay 
within the collinear factorization framework in what follows.  

The theoretical updates implemented 
%new results reported 
in this work are the following:
\begin{itemize}
\item{} The $c$-quark contribution to the coefficient function of the two-gluon DA;
\item{} Complete NLO treatment of the scale-dependence of DAs including quark-gluon mixing; 
\item{} Consistent treatment of the corrections due to the strange quark mass to $\mathcal{O}(m_s)$
        accuracy including an update of the $SU(3)$-breaking corrections in twist-four DAs;
\item{} Partial account of the anomalous contributions and implementation of $\eta-\eta'$ mixing schemes in the twist-four DAs.
\end{itemize}
We further use these improvements for a numerical analysis of the existing space-like and time-like data, 
including a careful analysis of the uncertainties, and the prospects to constrain the two-gluon $\eta^{(')}$ DAs 
if more precise data on transition FFs become available.

The presentation is organized as follows.
Section \ref{sec:DA} is introductory. We collect here the definitions 
for twist 2 and 3 DAs and introduce necessary notation in both the quark-flavor and singlet-octet bases. 
Different mixing schemes are introduced and discussed.
Section~\ref{sec:pQCD} is devoted to the calculation 
of the $\gamma^*\gamma \to \eta, \eta'$ electromagnetic transition FFs in the 
collinear factorization framework. Complete NLO expressions for the leading twist 
contributions are given. We also demonstrate the cancellation of the 
end point divergences in twist-four contributions at the tree (LO) level. 
The necessity to distinguish between the notion of ``power suppressed'' and ``higher-twist''
contributions is emphasized. A separate subsection contains the discussion of the 
difference of time-like and space-like FFs in pQCD; the results are compared to 
data~\cite{Aubert:2006cy}.   
In Section~\ref{sec:LCSR}  we start by explaining why the twist expansion of the product 
of electromagnetic currents does not provide the complete result for the FFs if one of the photons
is real, and present the calculation of the remaining soft contributions 
within the LCSR framework that is based on dispersion relations and quark-hadron duality.
A detailed numerical analysis of the space-like experimental data in this framework 
is presented in Section~\ref{sec:NUMERICS}. The final Section~\ref{sec:SUMMARY} is reserved 
for a summary and outlook.

The paper contains two appendices where more technical material and/or long expressions 
are collected.
Appendix~\ref{App:HT} is devoted to the two- and three-particle twist-four DAs of the $\eta, \eta'$ mesons. 
It contains an update of the existing expressions \cite{Braun:1989iv,Ball:1998je,Ball:2006wn}
taking into account $SU(3)$ breaking effects, and also a partial calculation
of anomalous contributions to the higher-twist DAs that arise from the axial anomaly.   
In Appendix \ref{App:RG} complete NLO expressions for the scale dependence 
of the leading-twist DAs are presented.

%%%%%%%%%%%%%%%%%%%%%%%%%%%%%%%%%%%%%%%%%%%%%%%%%%%%%%%%%%%%%%%%%%%%%%%%%%%%%%%%%%%%%%%%%%%%%%%%%%%

%%%%%%%%%%%%%%%%%%%%%%%%%%%%%%%%%%%%%%%%%%%%%%%%%%%%%%%%%%%%%%%%%%%%%%%%%%%%%%%%%%%%%%%%%%%%%%%%%%%
%
\section{$\eta$, $\eta'$ mixing and distribution amplitudes}
\label{sec:DA}
%
%%%%%%%%%%%%%%%%%%%%%%%%%%%%%%%%%%%%%%%%%%%%%%%%%%%%%%%%%%%%%%%%%%%%%%%%%%%%%%%%%%%%%%%%%%%%%%%%%%%

The description of the transition FFs $\gamma^*\gamma \to \eta,\eta'$ requires knowledge of the momentum fraction
distributions of valence quarks in the mesons at small transverse separations, the meson distribution 
amplitudes. We define the leading twist DA for a given quark flavor at a given scale $\mu$ as
\begin{eqnarray}
\lefteqn{\langle 0 |\bar q (z_2 n) \slashed{n}\gamma_5 q(z_1n) |M(p)\rangle = }
\nonumber\\&&{}\hspace*{0.5cm} 
= i F^{(q)}_M  (pn) \int_0^1 du\, e^{-i z_{21}^u (pn)} \phi_M^{(q)}(u,\mu)\,,  
\nonumber\\
\lefteqn{\langle 0 |\bar s (z_2 n) \slashed{n}\gamma_5 s(z_1n) |M(p)\rangle = }
\nonumber\\&&{}\hspace*{0.5cm} 
= i F^{(s)}_M (pn) \int_0^1 du\, e^{-i z_{21}^u (pn)} \phi_M^{(s)}(u,\mu)\,, 
\label{eq:DA-LT}
\end{eqnarray} 
where $q=u$ or $d$, $n_\mu$ is an auxiliary light-like vector, $n^2=0$,  and
we use a notation 
\begin{equation}
 z_{21}^u = \bar u z_2 + u z_1\,,\qquad \bar u = 1-u\,.
\end{equation}
In the following we also abbreviate
\begin{equation}
     z_{21} = z_2-z_1\,.
\end{equation}
The gauge links between the quark fields are implied.
In all equations $M=\eta,\eta'$ denotes the physical pseudoscalar meson state. 
We assume exact isospin symmetry and identify 
\begin{align}
  m_q = \frac12 (m_u+m_d)\,.
\end{align}
The normalization is chosen such that
\begin{equation}
 \int_0^1 du\, \phi_{M}^{(q,s)}(u,\mu) = 1\, 
\end{equation}
and the couplings $F^{(u)}_M = F^{(d)}_M$, $F^{(s)}_M$ are the matrix elements of flavor-diagonal axial vector
currents which we also write in the form
\begin{align}
   F^{(u)}_M =  F^{(d)}_M =  \frac{f^{(q)}_M}{\sqrt{2}}\,, &&   F^{(s)}_M =  f^{(s)}_M\,,    
\label{eq:bigF}
\end{align}
where
\begin{align}
 \langle 0| J^{(r)}_{\mu5}|M(p)\rangle = i f^{(r)}_M p_\mu\,, &&  r = q,s\,, 
\end{align}   
with the currents
 \begin{align}
 J^{(q)}_{\mu5} =\frac{1}{\sqrt{2}}\Big[\bar u \gamma_\mu\gamma_5 u + \bar d \gamma_\mu\gamma_5 d\Big],
 && 
 J^{(s)}_{\mu5} = \bar s \gamma_\mu\gamma_5 s\,.
\end{align} 
The scale dependence of the DAs can be simplified by introducing flavor-singlet and flavor-octet
combinations
\begin{align}
 f_M^{(8)}\phi^{(8)}_M &= \sqrt{\frac13} f_M^{(q)}\phi^{(q)}_M - \sqrt{\frac23} f_M^{(s)}\phi^{(s)}_M \,,
\notag\\
 f_M^{(1)}\phi^{(1)}_M &= \sqrt{\frac23} f_M^{(q)}\phi^{(q)}_M  +  \sqrt{\frac13} f_M^{(s)}\phi^{(s)}_M\,.  
\label{eq:SOQF}
\end{align} 
Here 
\begin{align}
 \langle 0| J^{(i)}_{\mu5}|M(p)\rangle = i f^{(i)}_M p_\mu\,, && i = 1,8\,,
\end{align}
where $J^{(1)}_{\mu5}$ and  $J^{(8)}_{\mu5}$ denote the $SU(3)$ flavor-singlet and octet currents
\begin{align}
 J^{(1)}_{\mu5} =& \frac{1}{\sqrt{3}}\Big[\bar u \gamma_\mu\gamma_5 u + \bar d \gamma_\mu\gamma_5 d + \bar s \gamma_\mu\gamma_5 s\Big],
\notag\\
 J^{(8)}_{\mu5} =& \frac{1}{\sqrt{6}}\Big[\bar u \gamma_\mu\gamma_5 u + \bar d \gamma_\mu\gamma_5 d -2 \bar s \gamma_\mu\gamma_5 s\Big].
\end{align}
Eq.~(\ref{eq:SOQF}) can be viewed as an orthogonal transformation from the quark-flavor (QF) to the singlet-octet (SO) basis
\begin{align}
\begin{pmatrix} f_M^{(8)}\phi^{(8)}_M(u,\mu) \\  f_M^{(1)}\phi^{(1)}_M (u,\mu)\end{pmatrix}   
&= U(\varphi_0) \begin{pmatrix} f_M^{(q)}\phi^{(q)}_M(u,\mu) \\  f_M^{(s)}\phi^{(s)}_M(u,\mu) \end{pmatrix}
\end{align}
where
\begin{align}
 U(\varphi_0) = 
 \begin{pmatrix}\sqrt{\frac13}& - \sqrt{\frac23} \\ \sqrt{\frac23} & \sqrt{\frac13} \end{pmatrix} 
 = \begin{pmatrix}\cos\varphi_0 & - \sin\varphi_0 \\ \sin\varphi_0 & \cos\varphi_0 \end{pmatrix}
\end{align}
with $\varphi_0 = \arctan(\sqrt{2})$. 

The main advantage of this representation is that the SO couplings and DAs do not mix with each other via renormalization.
In particular the octet coupling $f_M^{(8)}$ is scale-independent whereas the singlet coupling  $f_M^{(1)}$ evolves 
due to the $U(1)$ anomaly~\cite{Kodaira:1979pa}:
\begin{align}
  \mu \frac{d}{d\mu} f^{(1)}_{M}(\mu) & = - 4 n_f \left(\frac{\alpha_s}{2\pi}\right)^2 f^{(1)}_{M} +\mathcal{O}(\alpha_s^3),
\end{align}
or
\begin{align}
f_{M}^{(1)}(\mu )& = f_{M}^{(1)}(\mu_{0})\Big\{ 1+\frac{2 n_f}{\pi \beta_{0}}\Big[\alpha_s(\mu)- \alpha_s(\mu_{0})\Big]\Big\}, 
\label{eq:anomaly}
\end{align}
where $n_f$ is the number of light quark flavors.

The DAs can be expanded in terms of orthogonal polynomials $C_{n}^{3/2}(2u-1)$ that are 
eigenfunctions of the one-loop flavor-nonsinglet evolution equation:
\begin{equation}
 \phi_{M}^{(1,8)}(u,\mu) = 6u\bar u \Big[1+\!\!\sum\limits_{n=2,4,\ldots}\!\! c_{n,M}^{(1,8)}(\mu)
C_{n}^{3/2}(2u-1)\Big].
\label{phiq}
\end{equation} 
The sum in Eq.~(\ref{phiq}) goes over polynomials of even dimension $n=2,4,\ldots$. This restriction is 
a consequence of $C$-parity that implies that quark-antiquark DAs are symmetric functions 
under the interchange of the quark momenta
\begin{equation}
 \phi_{M}^{(1,8)}(u,\mu) = \phi_{M}^{(1,8)}(\bar u,\mu)\,. 
\label{eq:qDAsym}
\end{equation}
In addition we introduce a two-gluon leading-twist DA
$\phi_M^{(g)}(u,\mu)$, 
\begin{eqnarray}
\lefteqn{\langle 0 |G_{n\xi}(z_2n) \widetilde G^{n\xi}(z_1n) |M(p)\rangle = }
\nonumber\\&&{}\hspace*{0.5cm} 
= \frac{C_F}{2\sqrt{3}}f^{(1)}_M (pn)^2 \int_0^1 du\, e^{-i z_{21}^u (pn)} \phi_M^{(g)}(u,\mu)\,,  
\end{eqnarray}
where $C_F=4/3$, $\widetilde G_{\mu\nu}$ is the dual gluon field strength tensor 
$\widetilde G_{\mu\nu} = (1/2)\epsilon_{\mu\nu\alpha\beta}G^{\alpha\beta}$ and $G_{n\xi} = G_{\mu\xi} n^\mu$. 
We use the conventions  $\gamma_5 = i\gamma^0\gamma^1\gamma^2\gamma^3$ and 
$\epsilon_{0123} = 1$, following~\cite{Bjorken:1965zz}.
The gluon DA is antisymmetric  
\begin{equation}
 \phi_{M}^{(g)}(u,\mu) = - \phi_{M}^{(g)}(\bar u,\mu)\,
\end{equation}
and can be expanded in a series of Gegenbauer polynomials $C_{n-1}^{5/2}(2u-1)$ of odd dimension
\begin{equation}
\phi_{M}^{(g)}(u,\mu)
= 30 u^2\bar u^2 \!\!\sum\limits_{n=2,4,\ldots}\!\! 
 c_{n,M}^{(g)}(\mu)\,C_{n-1}^{5/2}(2u-1)\,.  
\label{phig}
\end{equation}
The flavor-octet Gegenbauer coefficients $c_{n,M}^{(8)}(\mu)$ are renormalized multiplicatively at LO, and
get mixed with the coefficients $c_{k,M}^{(8)}(\mu)$ with $k < n$ starting at NLO. 
The flavor-singlet coefficients $c_{n,M}^{(1)}(\mu)$  get mixed with the gluon coefficients $c_{n,M}^{(g)}(\mu)$ already at  LO,
and also with the coefficients of the polynomials with lower dimension starting at NLO, see Appendix~\ref{App:RG} for details.  
In what follows we refer to these coefficients as shape parameters. The values of shape parameters at a certain scale $\mu_0$
encode all nonperturbative information on the DAs.

In the exact $SU(3)$ flavor symmetry limit the $\eta$ meson is part of a flavor--octet, $\eta = \eta_8$, and $\eta'$ is a flavor--singlet,
$\eta' = \eta_1$. In this limit $f_\eta^{(s)} = -\sqrt{2} f^{(q)}_\eta $, $f_{\eta'}^{(s)} = 1/\sqrt{2} f^{(q)}_{\eta'}$ and $f^{(q)}_\eta = f_\pi$ where 
$f_\pi$ is the pion decay constant; in our normalization $f_\pi = 131$~MeV.  However, it is known empirically that the  
$SU(3)$-breaking corrections are large and have a rather nontrivial structure.  
In chiral effective theory the $\eta'$ meson can be included in the framework of the $1/N_c$ expansion~\cite{largeN}. 
In this approach the leading effect is due to the axial anomaly which introduces an effective mass term for the $\eta,\eta'$ states
that is not diagonal in the SO basis if $SU(3)$ flavor symmetry is broken. In addition, there is also an off-diagonal contribution to the 
kinetic term $\partial_\mu \eta_8 \partial^\mu \eta_1$ at loop level~\cite{Leutwyler:1997yr}. As a result, the relation of 
physical $\eta, \eta'$ states to the basic octet and singlet fields in the chiral Lagrangian, $\eta_8$ and $\eta_1$, becomes complicated and 
involves two different mixing angles, see, e.g., a discussion in Ref.~\cite{Feldmann:1998vh}. There is no reason to expect that these
mixing angles are the same for the matrix elements of all operators of higher dimension that determine moments of DAs. 
Thus the classification based on the SO mixing scheme without additional assumptions 
does not seem to be particularly useful in this context as the number of parameters is not reduced.
  
In the last years a specific approximation has become popular that we will refer to as 
the Feldmann--Kroll--Stech (FKS) scheme~\cite{Feldmann:1998vh}.  
This construction is motivated by the observation that the vector mesons $\omega$ and $\phi$ are 
to a very good approximation pure $\bar u u + \bar d d$ and $\bar s s$ states and the same pattern
is observed in tensor mesons. The smallness of mixing is a manifestation of the celebrated OZI rule that is 
phenomenologically very successful. If the axial $U(1)$ anomaly is the \emph{only} effect that 
makes the situation in pseudoscalar channels different, it is natural to assume that 
physical states are related to the flavor states by an orthogonal transformation
\begin{align}
\begin{pmatrix}|\eta\rangle \\ |\eta'\rangle  \end{pmatrix} =
U(\varphi)
\begin{pmatrix}|\eta_q\rangle \\ |\eta_s\rangle \end{pmatrix},
&&
U(\varphi) = \begin{pmatrix} \cos \varphi & - \sin \varphi\\ \sin \varphi & \cos\varphi \end{pmatrix}.
\end{align}
This \emph{state} mixing is a very strong assumption that implies that the same mixing pattern 
applies to the decay constants and, more generally, to the wave functions so that 
\begin{align}
 \begin{pmatrix} f_\eta^{(q)} & f_\eta^{(s)} \\ f_{\eta'}^{(q)} & f_{\eta'}^{(s)} \end{pmatrix}
 =& U(\varphi)
%\begin{pmatrix} \cos \varphi & - \sin \varphi \\ \sin \varphi  & \cos\varphi \end{pmatrix}
     \begin{pmatrix} f_q & 0 \\ 0 & f_s \end{pmatrix}. 
\label{eq:QF}
\end{align} 
and    
\begin{align}
 \begin{pmatrix} f_\eta^{(q)}\phi_\eta^{(q)} & f_\eta^{(s)}\phi_\eta^{(s)} \\ f_{\eta'}^{(q)} \phi_{\eta'}^{(q)} & f_{\eta'}^{(s)}\phi_{\eta'}^{(q)} \end{pmatrix}
 = U(\varphi)
 %\begin{pmatrix} \cos \varphi& - \sin \varphi\\ \sin \varphi& \cos\varphi\end{pmatrix}
     \begin{pmatrix} f_q\phi_q & 0 \\ 0 & f_s\phi_s \end{pmatrix}. 
\label{eq:QFt2}
\end{align}
with the same mixing angle $\varphi$.
 
This is a far reaching conjecture that allows one to reduce the four DAs of the physical 
states $\eta,\eta'$ to the two DAs  $\phi_q(u,\mu)$, $\phi_s(u,\mu)$ of the flavor states:
\begin{align}
&\phi_\eta^{(q)}(u) =  \phi_{\eta'}^{(q)}(u)  = \phi_q(u)\,,   
\notag\\
&\phi_\eta^{(s)}(u) =  \phi_{\eta'}^{(s)}(u)  = \phi_s(u)\,. 
\label{eq:QFmodel1}
\end{align}
The singlet and octet DAs in this scheme are given by
\begin{align}
 \begin{pmatrix}
  f_\eta^{(8)}\phi_\eta^{(8)} & f_\eta^{(1)}\phi_\eta^{(1)} \\ f_{\eta'}^{(8)}\phi_{\eta'}^{(8)} & f_{\eta'}^{(1)}\phi_{\eta'}^{(1)} 
 \end{pmatrix}
=&
 U(\varphi) \begin{pmatrix} f_q\phi_q & 0 \\ 0 & f_s\phi_s \end{pmatrix} U^T(\varphi_0) 
\label{relation}
\end{align}
and the same relation is valid separately for the couplings $f^{(r)}_M$ and the couplings 
multiplied by the shape parameters $f^{(r)}_M c^{(r)}_{n,M}$ (\ref{phiq}).
The couplings $f_q, f_s$ and the mixing angle $\phi$ in the FKS scheme have been determined in
Ref.~\cite{Feldmann:1998vh} from a fit to experimental data.
\begin{align}
f_q =& (1.07\pm 0.02)f_\pi\,, 
\notag\\ 
f_s =& (1.34\pm 0.06)f_\pi\,, 
\notag\\ 
\varphi =& 39.3^{\circ}\pm 1.0^\circ\,.
\label{FKSvalues-old}
\end{align}
A newer analysis~\cite{Escribano:2005qq} exploiting more recent data but only a subset of the processes investigated
in~\cite{Feldmann:1998vh} yields 
\begin{align}
f_q =& (1.09\pm 0.03)f_\pi\,, 
\notag\\ 
f_s =& (1.66\pm 0.06)f_\pi\,, 
\notag\\ 
\varphi =& 40.7^{\circ}\pm 1.4^\circ\,.
\label{FKSvalues-new}
\end{align}
where the mixing angle is the average of $\varphi_q$ and $\varphi_s$ from~\cite{Escribano:2005qq}.
The difference between the two sets can be viewed as an intrinsic uncertainty of the FKS approximation.  
For consistency with earlier work, e.g.~\cite{Kroll:2013iwa}, we will accept by default the
original set of parameters from Ref.~\cite{Feldmann:1998vh}, Eq.~(\ref{FKSvalues-old}), for numerical calculations in this work. 
A recent discussion of the the ongoing investigations of $\eta-\eta'$ mixing 
from a more general perspective can be found in~\cite{DiDonato:2011kr}.

\begin{figure}[t]
\begin{center}
\includegraphics[width=.44\textwidth,clip=true]{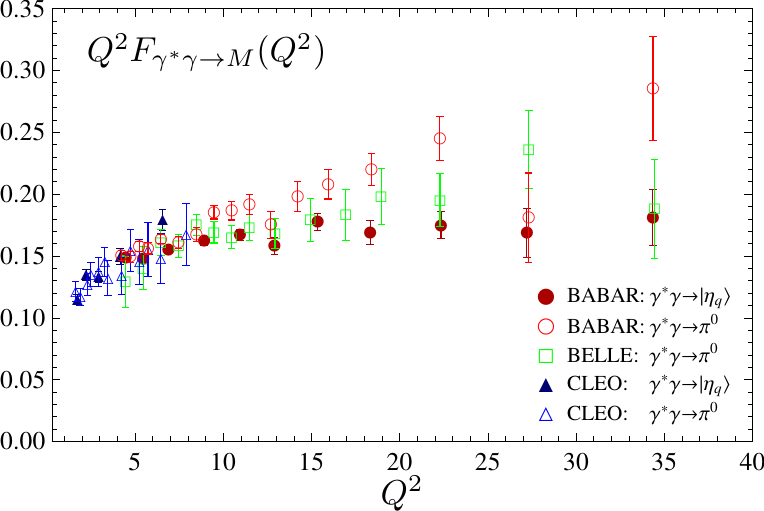}
\end{center}
\caption{The experimental data on $\gamma^*\gamma\to \pi^0$~\cite{Aubert:2009mc,Uehara:2012ag,Gronberg:1997fj} 
(open symbols) compared with the non-strange component of the eta meson transition FF, $\gamma^*\gamma\to |\eta_q\rangle$, 
(filled symbols), from the combination of BaBar and CLEO measurements~\cite{BABAR:2011ad,Gronberg:1997fj} 
on $\eta$ and $\eta'$ production in the FKS mixing scheme, Eqs.~(\ref{eq:QF}), (\ref{eq:QFt2}).
}
\label{fig:FKS-Data}
\end{figure}

Since the flavor-singlet and flavor-octet couplings have different scale dependence, Eq.~(\ref{relation}) 
cannot hold at all scales. It is natural to assume that the FKS scheme refers to a low renormalization scale $\mu_0 \sim 1$~GeV
and the DAs at higher scales are obtained by QCD evolution (that also generates nonvanishing OZI-violating contributions).
Figure~\ref{fig:FKS-Data} shows a comparison of the $\gamma^*\gamma\to \pi^0$ experimental data with the
non-strange $\gamma^*\gamma \to \vert \eta_q\rangle$ FF extracted from the combination of BaBar and CLEO measurements of 
$\gamma^*\gamma\to \eta$ and  $\gamma^*\gamma\to\eta'$ assuming the FKS mixing scheme. Were this scheme exact, the two FFs 
would coincide in the whole $Q^2$ range, up to tiny isospin breaking corrections.
It is seen that the existing measurements do not contradict the FKS approximation at low-to-moderate $Q^2 \lesssim 10$~GeV$^2$,
whereas at larger virtualities the comparison is inconclusive because of significant discrepancies between the BaBar and Belle
pion data.
The BaBar data taken alone show a dramatic difference between the $\gamma^*\gamma\to\pi^0$ and $\gamma^*\gamma\to |\eta_q\rangle$
FFs at large virtualities which cannot be explained by perturbative evolution effects. If this difference were confirmed, 
it would be a stark indication that the concept of state mixing is not applicable to the $\eta$ and $\eta'$ DAs so that the 
corresponding relations between higher-order Gegenbauer coefficients are strongly broken already at a low scale.  

Staying with the state mixing picture, 
for the gluon DA we have to assume that 
$$\langle 0 |G_{n\xi}(z_2n) \widetilde G^{n\xi}(z_1n) |\eta_q\rangle = \langle 0 |G_{n\xi}(z_2n) \widetilde G^{n\xi}(z_1n) |\eta_s\rangle$$
and as a consequence
\begin{align}
 \phi_\eta^{(g)}(u) =  \phi_{\eta'}^{(g)}(u)\,,    
\label{eq:QFmodel2}
\end{align}
that is similar to Eq.~(\ref{eq:QFmodel1}).

Two-particle twist-three DAs for the strange quarks can be defined as
\begin{equation}
2m_s \langle 0 | \bar{s}(z_2 n) i\gamma_5 s(z_1 n) |M(p)\rangle 
=
\int_{0}^{1}\!\!du\, e^{-iz_{21}^{u}(pn)} \phi_{3M}^{(s)p}(u)\,, 
\label{TW3:ps}
\end{equation}
and
\begin{eqnarray}
\lefteqn{
2m_s \langle 0 | \bar{s}(z_2 n) \sigma_{\mu\nu} \gamma_5 s(z_1 n) |M(p)\rangle=}
\nonumber\\&=& 
\frac{iz_{12}}{6}(p_\mu n_\nu-p_\nu n_\mu) \int_0^1 \!\!du\, e^{-iz_{21}^{u}(pn)} \phi_{3M}^{(s)\sigma}(u)
%\nonumber\\ 
\label{TW3:si}
\end{eqnarray}
with the normalization condition
\begin{equation}
 \int_0^1 du\, \phi_{3M}^{(s)p}(u) =  \int_0^1 du\, \phi_{3M}^{(s)\sigma}(u) =  H_{M}^{(s)}\,, 
\label{norm3}
\end{equation}
where
\begin{align}
H_{M}^{(s)} &= m_{M}^2 F_{M}^{(s)} - a_{M}\,,
\notag\\ 
  a_M &= \langle 0 | \frac{\alpha_s}{4\pi}G^A_{\mu\nu}\widetilde G^{A,\mu\nu}|M(p)\rangle\,, 
\label{eq:aM}
\end{align}
that follows from the anomaly relation 
\begin{equation}
  \partial^\mu  J^{(s)}_{\mu5} =  2m_s \bar s i\gamma_5 s + \frac{\alpha_s}{4\pi}G^A_{\mu\nu}\widetilde G^{A,\mu\nu}. 
\label{anomaly}
\end{equation}
Twist-three DAs for the light $q=u,d$ quarks can be defined by similar expressions with obvious 
substitutions $s\to q$, e.g. $H_{M}^{(q)} = m_{M}^2 F_{M}^{(q)} - a_{M}$.
In what follows we also use the notation, cf.~(\ref{eq:bigF}),
\begin{align}
   H^{(u)}_M =  H^{(d)}_M =  \frac{h^{(q)}_M}{\sqrt{2}}\,, &&   H^{(s)}_M =  h^{(s)}_M\,.    
\label{eq:bigH}
\end{align}
%$f_{M}^{(s)} \to f_{M}^{(q)}/\sqrt{2}$ and $h_{M}^{(s)} \to h_{M}^{(q)}/\sqrt{2}$. 
We do not present here the definitions of three-particle quark-antiquark-gluon twist-three DAs as it turns out that 
they do not contribute to the FFs of interest at LO  in perturbation theory. 

Assuming the FKS mixing scheme at low scales one can rewrite the four DAs 
$\phi_{3M}^{(q,s) p }$ in terms of two functions as in Eq.~(\ref{eq:QFt2}),
and similar for $\phi_{3M}^{(q,s)\sigma}$, introducing two new parameters $h_q$ and $h_s$~\cite{Beneke:2002jn}
\begin{align}
  h_q = 0.0015\pm 0.004\,, && h_s = 0.087\pm 0.006\,.
\label{hvalues}
\end{align}
Note that $h_q$ is small and consistent with zero. It is easy to convince oneself that matrix elements
of operators with even number of $\gamma$-matrices enter the calculation of 
the $\gamma^*\gamma\to\eta$  and $\gamma^*\gamma\to\eta'$  transition FFs always multiplied 
by quark masses, as on the left-hand-side (l.h.s.) of Eqs.~(\ref{TW3:ps}), (\ref{TW3:si}).
In this situation the contribution of light $q=u,d$ quarks is tiny and can safely be neglected.
To this accuracy
\begin{align}
  \phi_{3\eta'}^{(s)p}(u) =  \cos\varphi \, \phi_{3s}^{p}(u)\,, 
&&  
  \phi_{3\eta}^{(s)p}(u) =    -\sin\varphi \, \phi_{3s}^{p}(u)\,, 
\notag\\
  \phi_{3\eta'}^{(s)\sigma}(u) =  \cos\varphi \, \phi_{3s}^{\sigma}(u)\,, 
&&  
  \phi_{3\eta}^{(s)\sigma}(u) =    -\sin\varphi \, \phi_{3s}^{\sigma}(u)\,, 
  \label{QFS:Tw3}
\end{align}
where
\begin{align}
 \phi_{3s}^{p}(u) &= h_s + 60 m_s f_{3s} C^{1/2}_2(2u-1) +\ldots\,,
\notag\\
 \phi_{3s}^{\sigma}(u) &= 6\bar u u \Big[h_s + 10  m_s f_{3s} C^{3/2}_2(2u-1) +\ldots \Big]. 
\end{align}
The ellipses stand for the contributions of higher conformal spin and corrections $\mathcal{O}(m_s^2)$ 
which we neglect for consistency with the calculation of twist-four corrections (see the next section).
The coupling $f_{3s}$ is defined as 
\begin{align}
\langle 0 |\bar s \sigma_{n\xi}\gamma_5 g G^{n \xi} s | \eta_s (p) \rangle &= 2i (pn)^2 f_{3s}  
\end{align}
and we assume that $f^{(s)}_{3\eta'} =  \cos\varphi \,f_{3s}$, $f^{(s)}_{3\eta} = - \sin\varphi \,f_{3s}$.
The corresponding coupling for the charged $\pi$ meson is estimated to be (at the scale 1 GeV)~\cite{Ball:2006wn}
\begin{equation}
  f_{3\pi} \sim 0.0045~\text{GeV}^2.
\end{equation} 
Lacking any information about the flavor-singlet contribution, we adopt this
number as a (possibly crude) estimate for $f_{3s}$. With this choice 
\begin{align}
      \frac{2m_s f_{3s}}{h_s} \sim 0.01
\end{align} 
and one may hope that the corresponding ambiguity in FF predictions is not very large.
We will return to this question in the next section.
The scale dependence of  $f_{3s}$ is given by~\cite{Ball:2006wn} 
\begin{align}
  f_{3s}(\mu) &= L^{55/(9\beta_0)} f_{3s}(\mu_0) + \mathcal{O}(m_s f_s)
%\frac{2}{19}\left(L^{4/\beta_0}- L^{55/(9\beta_0)}\right)(m_s f_s)(\mu_0)  
\end{align} 
where $L=\alpha_s(\mu)/\alpha_s(\mu_0)$.
%where $\beta_0 = 11-(2/3)n_f$. Note that even if this coupling vanishes at a low scale, it will be
%generated at higher scales through mixing with the leading twist contributions.

Finally, we will need the DAs of twist-four that are rather numerous.
The corresponding expressions, including some new results, are collected in Appendix~\ref{App:HT}.

%%%%%%%%%%%%%%%%%%%%%%%%%%%%%%%%%%%%%%%%%%%%%%%%%%%%%%%%%%%%%%%%%%%%%%%%%%%%%%%%%%%%%%%%%%%%%%%%%%%
%
\section{$\gamma^*\gamma\to\eta,\eta'$  form factors in  QCD factorization}
\label{sec:pQCD}
%
%%%%%%%%%%%%%%%%%%%%%%%%%%%%%%%%%%%%%%%%%%%%%%%%%%%%%%%%%%%%%%%%%%%%%%%%%%%%%%%%%%%%%%%%%%%%%%%%%%%

%%%%%%%%%%%%%%%%%%%%%%%%%%%%%%%%%%%%%%%%%%%%%%%%%%%%%%%%%%%%%%%%%%%%%%%%%%%%%%%%%%%%%%%%%%%%%%%%%%%
%
\subsection{Leading twist}
\label{sec:LT}
%
%%%%%%%%%%%%%%%%%%%%%%%%%%%%%%%%%%%%%%%%%%%%%%%%%%%%%%%%%%%%%%%%%%%%%%%%%%%%%%%%%%%%%%%%%%%%%%%%%%%

The FFs $F_{\gamma^*\gamma^* \to M}(q_1^2,q_2^2)$,  $M=\eta,\eta'$
describing the meson transition in two (in general virtual) photons are defined by the following matrix
element of the product of two electromagnetic currents
\begin{eqnarray}
\lefteqn{\hspace*{-0.5cm}
 \int d^4 x\, e^{iq_1x}\,\langle M (p)|T\{j^{\rm em}_\mu(x)j^{\rm em}_\nu(0)\}|0\rangle =}
\nonumber\\
&&\hspace*{1.5cm}=\,  i e^2 \varepsilon_{\mu\nu\alpha\beta} q_1^\alpha q_2^\beta F_{\gamma^*\gamma^* \to M}(q_1^2,q_2^2)\,,
\label{eq:Fgamma}
\end{eqnarray}
where
$$j^{\rm em}_\mu(x) = e_u\bar u(x)\gamma_\mu u(x) +  e_d\bar d(x)\gamma_\mu d(x)+\ldots,$$
$p$ is the meson momentum and $q_2 = q_1 +p$. We will mainly consider the space-like FF,
in which case photon virtualities are negative.
In the experimentally relevant situation one virtuality is large
and the second one small (or zero). For definiteness we take
\begin{equation}
 q_1^2 = -Q^2\,,\qquad q_2^2 = -q^2\,,
\end{equation}
assuming that $q^2 \ll Q^2$. Most of the following equations are written for $q^2=0$,
and we use a shorthand notation
$$F_{\gamma^*\gamma \to M}(Q^2)\equiv F_{\gamma^*\gamma^* \to M}(q_1^2 = -Q^2,q^2=0).$$

The leading contribution ${\mathcal O}(1/Q^2)$ to the FFs can be written in factorized form
as a convolution of leading-twist DAs with coefficient functions that can be calculated in QCD perturbation
theory. 

The contribution of heavy (charm) quarks requires some attention. 
There are two basic possibilities to take into account heavy quarks in the QCD factorization 
formalism~\cite{Witten:1975bh,Collins:1978wz,Collins:1986mp,Collins:1998rz}
 which correspond, essentially, to the two choices of the (physical) factorization scale. 
It can be smaller, $\mu \ll m_h$, or larger, $\mu \gg m_h$
than the heavy quark mass.
 If  $\Lambda_{\rm QCD} \ll \mu  \ll m_h, Q$, i.e. if the (heavy) quark mass $m_h$ is very large, of the order of the photon 
virtuality $m_h\sim Q$, it is natural to write the structure function as a convolution of coefficient functions and
parton densities that involve only light quark flavors $u,d,s$ and gluons.
This approach is usually referred to as the decoupling scheme, or fixed flavor number scheme (FFNS).
Another possibility is to assume the hierarchy $\Lambda_{\rm QCD}, m_h \ll \mu \ll Q$
(which implies $m_h \ll Q$) and write the FFs as sum involving heavy flavors. This is
usually dubbed variable flavor number scheme (VFNS), with $\overline{\text{MS}}$ subtraction
for all flavors. 

In this work we adopt the first scheme which has the advantage that the complete heavy quark dependence 
is retained in the coefficient functions. A potential problem in this case is that for $m_h \ll Q$ 
the coefficient functions involve large logarithms $\sim \ln Q^2/m_h^2$ which one would like to resum
to all orders. This resummation is naturally done in the VFNS schemes where it corresponds
to the resummation of collinear logarithms using the ERBL equation, but the price to pay
is that this can only be done to leading power  accuracy in the $m_h^2/Q^2$ expansion.
There exists a vast literature devoted to heavy quark contributions to deep inelastic lepton 
hadron scattering (DIS), discussing how the advantages of both approaches can be combined by 
matching at the scale $\mu\simeq m_h$, see e.g.~\cite{Collins:1986mp}. 
We leave such improvements for future work, as the numerical impact of resummation on the transition 
FFs is not likely to be large. For the same reason we do not take into account
terms $\sim\alpha_s^2 \ln Q/m_h$ in the coefficient functions of light quark DAs. 

Thus we write
\begin{align}
  F_{\gamma^*\gamma \to M}(Q^2) =&  \frac{f_M^{(8)}}{3\sqrt{6}}\!\int_{0}^{1}\! du\, T^{(8)}_H(u,Q^2,\mu,\alpha_s(\mu))\phi^{(8)}_M(u,\mu)
\notag\\
                        +&  \frac{2f_M^{(1)}}{3\sqrt{3}}\!\int_{0}^{1}\! du\, T^{(1)}_H(u,Q^2,\mu,\alpha_s(\mu))\phi^{(1)}_M(u,\mu) 
\notag\\
                        +&  \frac{2f_M^{(1)}}{3\sqrt{3}}\!\int_{0}^{1}\! du\, T^{(g)}_H(u,Q^2,\mu,\alpha_s(\mu))\phi^{(g)}_M(u,\mu)\,, 
\label{eq:lt}
\end{align}
where $\phi^{(8,1,g)}_M(u,\mu)$ are the light quark octet (singlet), and gluon DAs defined in the previous Section.

\begin{figure}[t]
\begin{center}
\includegraphics[width=.45\textwidth,clip=true]{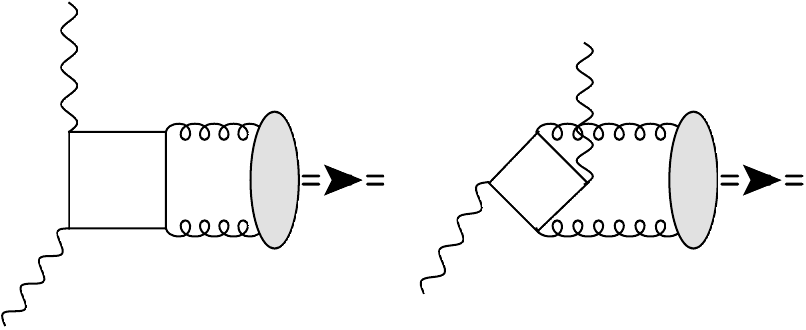}
\end{center}
\caption{Box diagrams contributing to the gluon coefficient function}
\label{fig:box}
\end{figure}

The coefficient function for the quark DA is known in the
$\overline{MS}$ scheme to NLO in the
strong coupling \cite{delAguila:1981nk,Braaten:1982yp,Kadantseva:1985kb}
and is the same for flavor-octet and flavor-singlet contributions.
Taking into account the symmetry of the quark DAs (\ref{eq:qDAsym}) it can be written as
\begin{eqnarray}
  T_H^{\rm NLO} &=& \frac{2}{u Q^2}
\biggl\{1+ C_F\frac{\alpha_s(\mu)}{2\pi}
\Big[\frac12 \ln^2 u -\frac12 \frac{u}{\bar u}\ln u
\nonumber\\
&&{} -\frac92 + \left(\frac32+\ln u \right)\ln\frac{Q^2}{\mu^2}
\Big]\biggr\}.
\label{eq:NLOcf}
\end{eqnarray}
The leading-order gluon coefficient function is calculated from the diagrams in Fig.~\ref{fig:box}.  
The contribution of light $u,d,s$ quarks reads~\cite{Kroll:2002nt,Kroll:2013iwa}
\begin{align}
 T_H^{g}\Big|_{\rm light} =& - 
C_F\frac{\alpha_s(\mu)}{2\pi}\frac{2\ln u }{\bar u^2 Q^2} \biggl\{\frac{1}{u} - 3  + \frac12 \ln u + \ln \frac{Q^2}{\mu^2}\biggr\}
\label{lightbox}
\end{align}
and the $c$-quark contribution is equal to %(this is probably a new result)
\begin{eqnarray}
T_H^{g}\Big|_{\rm charm}
 &=& C_F\frac{\alpha_s(\mu)}{2\pi}\frac23\frac1{u\bar u^2 Q^2}
\biggl\{ \ln^2\left[\frac{\beta(Q^2)+1}{\beta(Q^2)-1}\right]
\notag\\&&{}\hspace*{-1.3cm}
-u\ln^2\left[\frac{\beta(uQ^2)+1}{\beta(uQ^2)-1}\right]
%\notag\\&&{}
- 4\beta(Q^2) \ln\left[\frac{\beta(Q^2)+1}{\beta(Q^2)-1}\right] 
\notag\\&&{}\hspace*{-1.3cm}
+2(3u-1)\beta(uQ^2) \ln\left[\frac{\beta(uQ^2)+1}{\beta(uQ^2)-1}\right] 
            \biggr\},
\label{eq:charm}
\end{eqnarray}
where
\begin{align}
  \beta(Q^2) =& \sqrt{1+ \frac{4m_c^2}{Q^2}}. 
\end{align}
In numerical calculations we use the value $m_c= 1.42$~GeV for the $c$-quark pole mass. 
The $b$-quark contribution is given by the same expression with an obvious replacement of the quark mass $m_c \to m_b$
and extra factor $1/4$ from the electric charge $e_c^2\to e_b^2$. It is very small for the whole experimentally accessible region 
$Q^2 \lesssim 100$~GeV$^2$ and can safely be neglected. 

In the formal $Q^2\to\infty$ limit the transition form factors have to approach their asymptotic values~\cite{footnote1}
\begin{eqnarray}
\lefteqn{ \lim_{Q^2\to\infty} Q^2 F_{\gamma^*\gamma\to M}(Q^2){=}}
\notag\\
&=&
 \sqrt{\frac23}\Big[ f_M^{(8)} + 2\sqrt{2}f_M^{(1)}(\mu_0)\Big(1-\frac{2n_f}{\pi\beta_0}\alpha_s(\mu_0)\Big)\Big].
\label{eq:finrem}
\end{eqnarray}
Note that the scale dependence of the flavor-singlet axial coupling (\ref{eq:anomaly}) gives rise to a finite renormalization
factor $\sim 0.85$ which is not negligible. Using $n_f=4$, $\mu_0=1$~GeV, $\alpha_s(1$~GeV$) = 0.5$ and the FKS parameters in~(\ref{FKSvalues-old}) we obtain 
\begin{align}
 Q^2 F^{\rm asy}_{\gamma^*\gamma\to \eta}(Q^2)  &\to 0.173\,  (0.158)~\text{GeV}\,, 
\notag\\
 Q^2 F^{\rm asy}_{\gamma^*\gamma\to \eta'}(Q^2) &\to 0.247\,  (0.270)~\text{GeV}\,.
\label{eq:asy}
\end{align}
The asymptotic FF values corresponding to the parameter set in~(\ref{FKSvalues-new}) are shown in parenthesis for comparison.  
The finite renormalization correction to the flavor-singlet contribution is not taken into account
in~\cite{BABAR:2011ad,Kroll:2002nt,Kroll:2013iwa}.
It is only a $\lesssim 5\%$ effect for the $\eta$-meson, but leads to
a 20\% reduction of the asymptotic value of the FF for the $\eta'$, in which case the effect is amplified by the cancellation 
between the flavor-singlet and flavor-octet contributions, 
$f^{(1)}_{\eta'} = 0.15\,(0.17)$, $f^{(8)}_{\eta'} = -0.06\,(-0.08)$. In this way the discrepancy between
the data~\cite{BABAR:2011ad} and the expected asymptotic behavior of the $\gamma^*\gamma\to \eta'$ FF is removed, see Section~\ref{sec:NUMERICS}.

%%%%%%%%%%%%%%%%%%%%%%%%%%%%%%%%%%%%%%%%%%%%%%%%%%%%%%%%%%%%%%%%%%%%%%%%%%%%%%%%%%%%%%%%%%%%%%%%%%%
%
\subsection{Higher twist corrections}
\label{sec:HT}
%
%%%%%%%%%%%%%%%%%%%%%%%%%%%%%%%%%%%%%%%%%%%%%%%%%%%%%%%%%%%%%%%%%%%%%%%%%%%%%%%%%%%%%%%%%%%%%%%%%%%

One source of power corrections $\sim 1/Q^2$ to the transition FFs 
$F_{\gamma^*\gamma^* \to M}$ corresponds to contributions of less singular terms $\sim 1/x^2$, $\ln x^2$, etc.,
as compared to the leading contribution $\sim 1/x^4$ in the operator product expansion of the two electromagnetic  currents 
in Eq.~(\ref{eq:Fgamma}). They can be calculated in terms of meson DAs 
of higher twist and will be referred  to as higher-twist corrections in what follows. 
To LO in perturbation theory one obtains including the twist-four contribution
\begin{eqnarray}
  Q^2 F_{\gamma^*\gamma \to M}(Q^2) &=&  2\!\! \sum\limits_{\psi=u,d,s}e^2_\psi F^{(\psi)}_M \int_0^1 \frac{du}{u} \bigg\{\phi_M^{(\psi)}(u)
\notag\\ &&{} - \frac{\bar u m_M^2}{Q^2} \phi_M^{(\psi)}(u) + \frac{1}{6u Q^2} \phi^{(\psi)\sigma}_{3M}(u)
\notag\\ &&{} - \frac{1}{uQ^2} \mathbb{A}^{(\psi)}_{4;M}(u)\biggr\} 
\end{eqnarray}
where the function $\mathbb{A}_{4;M}(u)$ is written in terms of two-particle and three-particle 
DAs of twist-four defined in Appendix~\ref{App:HT}:
\begin{eqnarray}
\mathbb{A}_{4;M}(u) &=& \frac14 \phi_{4M}(u) -\int_0^u\! d\alpha_1 \!\int_0^{\bar u}\! d\alpha_2 
\Big[\frac{1}{\alpha_3}\widetilde{\Phi}_{4M}(\underline{\alpha})
\nonumber\\&&{}
+ \frac{2u-1-\alpha_1+\alpha_2}{\alpha_3^2}\Phi_{4M}(\underline{\alpha})\Big]\Big|_{\alpha_3=1-\alpha_1-\alpha_2} 
\end{eqnarray}
Using explicit expressions for the twist-four DAs, see Appendix~\ref{App:HT}, we obtain
\begin{eqnarray}
\lefteqn{Q^2 F_{\gamma^*\gamma \to M}(Q^2)=}
\nonumber\\ &=&  2\!\!\! \sum\limits_{\psi=u,d,s}
\!\!\! e^2_\psi F^{(\psi)}_M
\biggl\{ 3(1+c^{(\psi)}_{2,M}) 
%\nonumber\\&&{}
- \frac{1}{Q^2}\biggl[\frac{h^{(\psi)}_M}{f^{(\psi)}_M} (2+3 c^{(\psi)}_{2,M})
\nonumber\\&&{}
+ \frac{80}{9}\delta^{2(\psi)}_M 
- \frac{h^{(\psi)}_M}{f^{(\psi)}_M}\biggl(\frac{67}{360}-\frac54 c^{(\psi)}_{2,M} \biggr)
%\nonumber\\&&{}\hspace*{5cm}
-\frac32 \frac{m_\psi f^{(\psi)}_{3M}}{f^{(\psi)}_M}\biggr]\biggr\},  
\nonumber\\
\end{eqnarray} 
where we included, for comparison, the leading-order leading twist contribution 
and ignored the scale dependence.
Note the following:
\begin{itemize}
\item{} The end-point divergence at $u\to 0$ in the contribution of the twist-three 
  DA $\phi_M^{(\psi)}(u)$ exactly cancels the similar divergence in the twist-four 
  contributions that are related to twist-three operators by equations of motion; 
  this cancellation is general and does not depend on the shape of the 
  twist-three DAs.
\item{} Assuming the FKS mixing scheme  the expression for the $1/Q^2$ correction
       (in square brackets) does not depend on the meson states, $\eta$ or $\eta'$.
        Using the numbers quoted in Eqs.~(\ref{FKSvalues-old}),(\ref{hvalues}) we
        obtain for the ratio
\begin{align}
         h^{(s)}_M/f^{(s)}_M & = 0.50 \pm 0.04~\text{GeV}^2,  
\end{align}
whereas the similar ratio for the light $u,d$ quarks is compatible with zero.          
\item{}  The higher-twist correction is dominated by the contribution of 
       $\delta^{2(\psi)}_M \simeq 0.2$~GeV$^2$ (see Appendix~\ref{App:HT}) 
       whereas the contribution of the twist-three quark-antiquark-gluon 
       matrix element $\sim m_s f^{(s)}_{3M}/f^{(s)}_M$ is completely negligible.
\end{itemize} 
Plugging in the numbers we obtain a rough estimate of the twist-four contribution  
\begin{align}
    F_{\gamma^*\gamma \to M}(Q^2) &=&  \biggl[1- \frac{0.9~\text{GeV}^2}{Q^2}\biggr] F^{\rm twist-2}_{\gamma^*\gamma \to M}(Q^2).  
\end{align}
This is a small correction. However, one can show that contributions of \emph{arbitrary} twist produce a $1/Q^2$ correction
as well (see a detailed discussion in \cite{Agaev:2010aq}), indicating that the light-cone dominance of the 
transition form factor with one virtual and one real photon does not hold beyond leading power accuracy.  
An estimate of the twist-six contribution~\cite{Agaev:2010aq} results in a small positive $1/Q^2$ correction,
enhanced by an additional $\ln Q^2$ factor. The mismatch of twist- and power-counting is due to the fact
that to power accuracy one must consider the contributions of large light-cone distances
between the currents, that are not ``seen'' in the twist expansion. To leading order in the QCD coupling
such terms can simply be added and there is no double counting. An example of such a correction is 
the contribution of real photon emission at large distances calculated in Ref.~\cite{Agaev:2010aq}:
\begin{eqnarray}
      F_{\gamma^*\gamma \to \pi^0}(Q^2) & = &  \frac{\sqrt{2}f_\pi}{3}\,
\frac{16\pi\alpha_s\chi \langle \bar q q\rangle^2 }{9 f_\pi^2 Q^4}
\nonumber\\&&{}\times \int_0^1 \!dx\,\frac{\phi^{p}_{3;\pi}(x)}{x}
\int_0^1 \!dy\,\frac{\phi_\gamma(y)}{\bar y^2}\,,
\label{eq:3b}
\end{eqnarray}

    %%%%%%%%%%
    %%%%%%%%%% Begin Table 1
    %%%%%%%%%%
\begin{table*}[t]
%[htbp]
\renewcommand{\arraystretch}{1.2}
\centering
\begin{tabular}{|c|l|l|l|l|| l|l|l|}\hline
 meson                  & scale      &  $f_2^{(q)}$     &  $f_2^{(s)}$     &  $f_2^{(g)}$      &  $f_4^{(q)}$     &  $f_4^{(s)}$     &  $f_4^{(g)}$ \\ \hline \hline
\multirow{2}{*}{$\eta$} & space-like & 0.126          & -0.037         & 0.010           & 0.105          & -0.030         & 0.006 \\
                        & time-like  & 0.113 + 0.032i & -0.033 - 0.009i& 0.011 - 0.001i  & 0.086 + 0.039i  & -0.025 - 0.011i& 0.006 + 0.001i        
\\  \hline\hline 
\multirow{2}{*}{$\eta'$}& space-like & 0.103          & 0.045          & 0.061           & 0.086          & 0.037          & 0.037  \\
                        & time-like  & 0.093 + 0.026i & 0.040 + 0.011i & 0.069 - 0.005i  & 0.070 + 0.032i & 0.030 + 0.014i & 0.040 + 0.005i         
\\  \hline\hline  
\end{tabular}
\caption{Coefficients (\ref{eq:time-space}) of the contributions of different Gegenbauer polynomials in the expansion of DAs to the 
transition form factors at the time-like $Q^2 = -s = -112$~GeV$^2$,
assuming validity of the FKS mixing scheme (\ref{FKSvalues-old}) at the low scale $\mu_0 =1$~GeV. 
The corresponding space-like coefficients for $Q^2 = 112$~GeV$^2$ are also given for comparison. All numbers in units of GeV.}
\label{tab:table1}
\renewcommand{\arraystretch}{1.0}
\end{table*}
    %%%%%%%%%%
    %%%%%%%%%% End Table 1
    %%%%%%%%%%

\noindent
where $\phi_\gamma(y)\simeq 6 y(1-y)$ is the leading-twist photon DA \cite{Balitsky:1989ry,Ball:2002ps}
and $\chi \simeq 3.5$~GeV$^{-2}$ (at the scale $\mu=1$~GeV) is the magnetic susceptibility of the quark 
condensate~\cite{Ioffe:1983ju,Belyaev:1984ic,Balitsky:1985aq,Ball:2002ps,Bali:2012jv}.
The integrals over the quark momentum fractions in (\ref{eq:3b}) are both logarithmically
divergent at the end-points $x\to0$, \mbox{$y\to1$}, which signals that there is an overlap with the
soft region.  Such soft contributions 
are related to the overlap between the light-cone wave functions of the 
pseudoscalar meson and the real photon and can be taken into account in the framework of LCSRs 
described in the next section.

%%%%%%%%%%%%%%%%%%%%%%%%%%%%%%%%%%%%%%%%%%%%%%%%%%%%%%%%%%%%%%%%%%%%%%%%%%%%%%%%%%%%%%%%%%%%%%%%%%%
%
\subsection{Time-like form factors}
\label{sec:timelike}
%
%%%%%%%%%%%%%%%%%%%%%%%%%%%%%%%%%%%%%%%%%%%%%%%%%%%%%%%%%%%%%%%%%%%%%%%%%%%%%%%%%%%%%%%%%%%%%%%%%%%

In Ref.~\cite{Aubert:2006cy} the processes $e^+ e^-\to\gamma^*\to(\eta,\eta')\gamma$ were studied at 
a center of mass energy of $\sqrt{s}=10.58$ GeV. The measurements can be interpreted in terms of the 
$\gamma^*\gamma\to \eta,\eta'$ FFs at remarkably high time-like photon virtuality $Q^2 = -s = -112$~GeV$^2$: 
\begin{align}
\vert Q^2 F_{\gamma^*\gamma\to\eta}(Q^2)\vert_{Q^2=-112\,\mbox{\scriptsize GeV}^2} &= (0.229\pm0.031)\,\text{GeV},
%\nonumber\\ &= (0.229\pm0.030\pm0.008) \,\mbox{GeV}^2, 
\nonumber\\
\vert Q^2 F_{\gamma^*\gamma\to\eta'}(Q^2)\vert_{Q^2=-112\,\mbox{\scriptsize GeV}^2}  &= (0.251\pm0.021)\,\text{GeV},
%\nonumber\\&=(0.251\pm0.019\pm0.008)\,\mbox{GeV}^2,
\label{eq:timedata}
\end{align}
where we added the statistical and systematic uncertainties in quadrature.
%the first uncertainty is statistical while the second is systematic.
Note that the time-like FFs are complex numbers, whereas only the absolute value is measured.

To leading twist accuracy, the time-like FFs can be obtained from their Euclidean (space-like) expressions 
by the analytic continuation
\begin{align}
  Q^2 \mapsto  - s - i\epsilon\,.
\end{align}
The imaginary parts arise both from the analytic continuation of the hard coefficient functions and the DAs which become 
complex at time-like scales $\mu^2\sim Q^2=-s$, see e.g.~\cite{Bakulev:2000uh}. 

Since transition form factors are linear functions of the meson DAs, the results of the QCD calculation can be written as
a sum of contributions of different Gegenbauer polynomials at the low reference scale
\begin{eqnarray}
\lefteqn{ \hspace*{-0.7cm} Q^2  F^{\rm twist-2}_{\gamma^*\gamma\to\eta}(Q^2)\big|_{Q^2=-112\,\mbox{\scriptsize GeV}^2}=}
\nonumber\\ &=& 
0.161~\text{GeV}  +
\sum_{p=q,s,g}\sum_{n=2,4,\ldots} f_{\eta;n}^{(p)}\, c_n^{(p)}(\mu_0^2)\,,
\nonumber\\
\lefteqn{ \hspace*{-0.7cm} Q^2 F^{\rm twist-2}_{\gamma^*\gamma\to\eta'}(Q^2)\big|_{Q^2=-112\,\mbox{\scriptsize GeV}^2}=}
\nonumber\\ &=& 
0.241~\text{GeV}  +
\sum_{p=q,s,g}\sum_{n=2,4,\ldots} f_{\eta';n}^{(p)}\, c_n^{(p)}(\mu_0^2)\,,
\label{eq:time-space}
\end{eqnarray}
where the asymptotic DA contributions are almost the same in the time-like and space-like regions, and the coefficients 
$f_{M;n}^{(p)} \equiv f_{M;n}^{(p)}(Q^2/\mu^2,\alpha_s(\mu^2);\mu_0^2)$ 
%$f_{\eta;n}^{(p)},f_{\eta';n}^{(p)}$
absorb all dependence on $Q^2$. Numerical values of these coefficients with the choice of factorization scale $\mu^2=Q^2$,
continued analytically to the time-like values $Q^2 = - s$, are presented for $\eta$ and $\eta'$ mesons in 
comparison with the corresponding space-like coefficients for $n=2,4$ in Table~1. 
Note that the Gegenbauer coefficients at the low scale $c_n^{(p)}(\mu_0)$ do not depend on the type of the meson --- $\eta$ or $\eta'$ ---
by assumption of the FKS state mixing. For this calculation we have taken the set of parameters in Eq.~(\ref{FKSvalues-old}).
The given numbers correspond to the choice of the scale $\mu^2=Q^2$, they change by at most 10\% if the scale is varied in the 
interval $Q^2/2 < \mu^2 < 2 Q^2$. 

We see that the coefficients of higher Gegenbauer polynomials are in general rather small, which is due to suppression by the 
anomalous dimensions. These coefficients acquire rather large phases, however, for realistic values of the Gegenbauer coefficients 
$c_{2,4}^{(q)}\sim c_{2,4}^{(s)}\approx0.1-0.2$ the corresponding contributions to the FF appear to be marginal as compared to the leading terms 
in~(\ref{eq:time-space}).
Thus the overall phase is small and the absolute values of the FF in the space-like and time-like regions remain close to each other.
This result is in agreement with the conclusion in~\cite{Bakulev:2000uh} that perturbative corrections cannot generate
a significant difference between the space-like and time-like transition FFs.    

Beyond the leading power accuracy the situation is less clear. Note that the overall $1/Q^2$ correction to the space-like transition form factors 
is negative (this can be shown in many ways, see, e.g.~\cite{Agaev:2010aq,Agaev:2012tm}) and by virtue of the sign change in $Q^2$ one expects a
positive correction to the time-like form factors if the analytic continuation is justified to power accuracy which is, however, not obvious.
The higher-twist contributions corresponding to less singular terms in the light-cone expansion of the product of the two electromagnetic currents 
are small and tend to have alternating signs, cf. the discussion in the previous section. They are unlikely to play any role at 
$|Q^2| \sim 100$~GeV$^2$. The soft contributions can, however, be significant.  

Within the LCSR approach to soft contributions discussed in the next section, their magnitude is correlated with the shape of the leading twist DA:
broader DAs generally lead to larger soft corrections and vice verse. A rough estimate (\ref{simple}) gives
\begin{align}
     Q^2 F_{\gamma^*\gamma\to\eta}(Q^2) \simeq  Q^2 F^{QCD}_{\gamma^*\gamma\to\eta}(Q^2)\left[1 - \frac{(3-7)~\text{GeV}^2}{Q^2}\right],  
     \label{eq:SoftEst}
\end{align}
where the larger number corresponds to a broad DA of the type~\cite{Agaev:2010aq} required to describe the BaBar data~\cite{Aubert:2009mc}
on $\gamma^*\gamma\to \pi^0$, and the smaller one is obtained for the asymptotic DA.
%a more conventional DA~\cite{Agaev:2012tm} that is sufficient e.g. to describe the BELLE measurement~\cite{Uehara:2012ag} of the same FF. 
Assuming that the soft correction  changes sign in the time-like region, we conclude 
that the difference between the time-like and space-like form factors at $|Q^2|=112$~GeV$^2$ can be of the order of  $\sim 5-13\%$ for the ``narrow'' 
and ``broad'' meson DA, respectively. This difference can further be enhanced by Sudakov-type corrections, see the 
discussion in~\cite{Bakulev:2000uh} and references therein. 

It is interesting that the experimental result for $\gamma^*\gamma\to\eta'$ at $Q^2=-112$\,GeV$^2$~\cite{Aubert:2006cy} is very close to the 
contribution of the asymptotic $\eta'$ meson DA in Eq.~(\ref{eq:time-space}), whereas the asymptotic contribution to $\gamma^*\gamma\to\eta$ is 
almost 50\% below the data, cf. (\ref{eq:timedata}). This result urgently needs verification. If correct, it can probably only be explained by 
much larger soft contributions alias a much broader DA of the $\eta$ meson as compared to $\eta'$, which would be in conflict with 
the state mixing approximation for DAs. 
 
%%%%%%%%%%%%%%%%%%%%%%%%%%%%%%%%%%%%%%%%%%%%%%%%%%%%%%%%%%%%%%%%%%%%%%%%%%%%%%%%%%%%%%%%%%%%%%%%%%%
%
\section{Light-Cone Sum Rules}
\label{sec:LCSR}
%
%%%%%%%%%%%%%%%%%%%%%%%%%%%%%%%%%%%%%%%%%%%%%%%%%%%%%%%%%%%%%%%%%%%%%%%%%%%%%%%%%%%%%%%%%%%%%%%%%%%

The LCSR approach was proposed in \cite{Balitsky:1986st,Balitsky:1989ry,Braun:1988qv,Chernyak:1990ag} 
and adapted for the present situation in \cite{Khodjamirian:1997tk}. This technique is well-known and has been
used repeatedly for 
$\gamma^*\gamma\to \pi^0$~\cite{Schmedding:1999ap,Bakulev:2001pa,Bakulev:2002uc,Bakulev:2003cs,Agaev:2005rc,Mikhailov:2009kf,Agaev:2010aq,Agaev:2012tm,Bakulev:2012nh,Stefanis:2012yw} so that in what follows we will only give a short introduction and present the necessary NLO expressions, 
generalized and/or adapted for the case of $\eta^{(')}$-mesons.      

The idea is to consider a more general transition FF for two nonvanishing photon virtualities, $q_1^2 = -Q^2$ and $q_2^2 = -q^2$,
and perform the analytic continuation to the real photon limit $q^2=0$ employing dispersion relations.

On the one hand, $F_{\gamma^*\gamma^* \to M}(Q^2,q^2)$
satisfies an unsubtracted dispersion relation in the variable $q^2$ for fixed $Q^2$.
Separating the contribution of the lowest-lying vector mesons $\rho,\omega$ we can write
\begin{eqnarray}
 F_{\gamma^*\gamma^* \to M}(Q^2,q^2) &=& \frac{\sqrt{2}f_\rho F_{\gamma^*\rho \to M}(Q^2)}{m^2_\rho+q^2}
\nonumber\\&&\hspace*{-1.5cm}{} +
\frac{1}{\pi}\int_{s_0}^\infty ds\,\frac{\mathrm{Im} F_{\gamma^*\gamma^* \to M}(Q^2,-s)}{s+q^2}\,,
\label{eq:DR}
\end{eqnarray}
where $s_0$ is some effective threshold. Here, the $\rho$ and $\omega$ contributions
are combined in one resonance term assuming $m_\rho\simeq m_\omega$ and the zero-width
approximation is used; $f_\rho\sim 200$~MeV is the usual vector meson decay constant.
Note that since there are no massless states, the real photon limit is recovered
by the simple substitution $q^2\to 0$ in (\ref{eq:DR}).

On the other hand, the same FF can be calculated using QCD perturbation theory and the OPE.
The QCD result obeys a similar dispersion relation
\begin{equation}
 F^{\rm QCD}_{\gamma^*\gamma^* \to M}(Q^2,q^2) = \frac{1}{\pi}\int_{0}^\infty ds\,\frac{\mathrm{Im} F^{\rm QCD}_{\gamma^*\gamma^* \to M}(Q^2,-s)}{s+q^2}\,.
\label{eq:DRQCD}
\end{equation}
The basic assumption, usually referred to as quark-hadron duality, is that the physical spectral density
above the threshold  $s>s_0$ coincides with the QCD spectral density as given by the OPE:
\begin{equation}
   \mathrm{Im}F_{\gamma^*\gamma^* \to M}(Q^2,-s) = \mathrm{Im}F^{QCD}_{\gamma^*\gamma^* \to M}(Q^2,-s).
\label{eq:duality}
\end{equation}
This equality has to be understood in the sense of distributions, with both 
sides integrated with a smooth test function. 

Equating the two representations in (\ref{eq:DR}) and (\ref{eq:DRQCD}) at $q^2\to-\infty$
and subtracting the contributions of $s>s_0$ from both sides one obtains
\begin{eqnarray}
  \sqrt{2}f_\rho F_{\gamma^*\rho \to M}(Q^2) = \frac{1}{\pi}\int_{0}^{s_0}\!\!ds\,
\mathrm{Im} F^{\rm QCD}_{\gamma^*\gamma^* \to M}(Q^2,-s)\,.
\end{eqnarray}
This relation explains why $s_0$ is usually referred to as the interval of duality.
The perturbative QCD spectral density $\mathrm{Im}F^{QCD}_{\gamma^*\gamma^* \to M}(Q^2,-s)$ is a
smooth function and does not vanish at small $s\to 0$. It is very different from the physical
spectral density $\mathrm{Im}F_{\gamma^*\gamma^* \to M}(Q^2,-s) \sim \delta(s-m_\rho^2)$.
However, the integral of the QCD spectral density over a certain region of energies coincides
with the integral of the physical spectral density over the same region; in this sense the QCD
description of correlation functions in terms of quark and gluons is dual to the description in terms of
hadronic states.

In practical applications of this method one uses a trick borrowed from QCD sum rules~\cite{SVZ}, 
to reduce the sensitivity to the duality assumption in Eq.~(\ref{eq:duality}) and also to suppress contributions arising from higher order
terms in the OPE. To this end one attempts to match the ``true'' and
calculated FF at a finite value $q^2 \sim 1-2$~GeV$^2$ instead of the $q^2\to\infty$
limit. This is done going over to the Borel representation $1/(s+q^2)\to \exp[-s/M^2]$
the final effect being the appearance of an additional weight factor under the integral
\begin{eqnarray}
  \sqrt{2}f_\rho F_{\gamma^*\rho \to M}(Q^2) &=& \frac{1}{\pi}\int_{0}^{s_0}ds\, e^{-(s-m^2_\rho)/M^2}\,
\nonumber \\&&{}\times
\mathrm{Im} F^{\rm QCD}_{\gamma^*\gamma^* \to M}(Q^2,-s)\,.
\label{eq:Frhogamma}
\end{eqnarray}
Varying the Borel parameter within a certain window one may test the
sensitivity of the results to a chosen model for the spectral density.

With this refinement, substituting Eq.~(\ref{eq:Frhogamma}) in (\ref{eq:DR}) and using
Eq.~(\ref{eq:duality}) we obtain for $q^2\to 0$
\begin{eqnarray}
  F^{\rm LCSR}_{\gamma^*\gamma \to M}(Q^2) &\!=\!&
\frac{1}{\pi}\!\int_{0}^{s_0}\!\!\! \frac{ds}{m_\rho^2}
\mathrm{Im} F^{\rm QCD}_{\gamma^*\gamma^* \to M}(Q^2\!,-s)
e^{(m^2_\rho-s)/M^2} \nonumber\\&&{} +
\frac{1}{\pi}\int_{s_0}^\infty \frac{ds}{s} \mathrm{Im} F^{\rm
QCD}_{\gamma^*\gamma^* \to M}(Q^2,-s)\,. 
\label{eq:22}
\end{eqnarray}
This expression contains two nonperturbative parameters, the vector meson mass $m_\rho^2$ and the effective threshold
$s_0\simeq 1.5$~GeV$^2$, as compared to the ``pure'' QCD calculations.

Taking into account Eq.~(\ref{eq:DRQCD}) one can rewrite the same result as
\begin{eqnarray}
  F^{\rm LCSR}_{\gamma^*\gamma \to M}(Q^2) &\!=\!& F^{\rm QCD}_{\gamma^*\gamma \to M}(Q^2)
\nonumber\\&&
{}\hspace*{-2cm}+ \frac{1}{\pi}\!\int_{0}^{s_0}\!\!\! \frac{ds}{m_\rho^2}
\Big[e^{(m^2_\rho-s)/M^2}- \frac{m_\rho^2}{s} \Big] \mathrm{Im} F^{\rm QCD}_{\gamma^*\gamma^* \to M}(Q^2\!,-s)\,,
\nonumber\\
\label{eq:23}
\end{eqnarray} 
separating the result of a ``pure'' QCD calculation and the correction.

To get an impression how this modification affects the QCD result,
we insert the leading order and leading twist expression for 
$\mathrm{Im} F^{\rm QCD}_{\gamma^*\gamma^* \to M}(Q^2\!,-s)$ and rewrite the dispersion integral in
terms of a variable $x = Q^2/(s+Q^2)$ that corresponds to the fraction of the meson momentum
carried by the interacting quark:
\begin{eqnarray}
F^{\rm LCSR}_{\gamma^*\gamma \to M}(Q^2) &\!=\!&\sum_{i=1,8} C^i f_M^{(i)}\frac{1}{Q^2}\left[\int_0^1\frac{dx}{\bar x}\phi_M^{(i)}(x)\right.
\nonumber\\
&+&\left.\int^1_{x_0}\!\!\frac{dx}{\bar x}\left(\frac{\bar x Q^2}{x m_\rho^2}e^{\frac{{x} m_\rho^2-\bar x Q^2}{{x}M^2}}- 1 \right)\phi_M^{(i)}(x)\right],
\nonumber\\
\end{eqnarray}
where $C^1=\frac{4}{3\sqrt{3}}$, $C^8=\frac{2}{3\sqrt{6}}$, $\bar x = 1-x$  and $x_0=\frac{Q^2}{s_0+Q^2}$. 
The first contribution is the LO perturbative result while the second part 
represents the soft end-point correction from the region $ x > x_0 = 1-\mathcal{O}(s_0/Q^2)$, due to the modification of the
spectral density in the LCSR framework.

For a rough estimate of the soft correction we expand the integrand for small $1-x_0$
\begin{eqnarray}
F^{\rm LCSR}_{\gamma^*\gamma \to M}(Q^2) &\!\approx\!&\sum_{i=1,8} C^i f_M^{(i)}\frac{1}{Q^2}\left[\int_0^1\frac{dx}{\bar x}\phi_M^{(i)}(x)\right.
\nonumber\\
&+&\left.\bar x_0\left(\frac{s_0}{2m_\rho^2} e^{\frac{m_\rho^2-s_0}{M^2}}-1\right)\phi_M^{'(i)}(0)\right],
\nonumber\\
\label{simple}
\end{eqnarray}
where $\phi_M^{'(i)}(0) \equiv (d/dx)\phi_M^{(i)}(x)|_{x=0}$ and
we assumed that the DA vanishes linearly at the end points. Using 
\begin{eqnarray}
\phi_M^{'(i)}(0)&=&3\biggl[2+\sum_{n=2,4,\ldots} (n+1)(n+2) c_n^{(i)}\biggr],
\nonumber\\
\int_0^1\!\frac{dx}{\bar x}\phi_M^{(i)}(x)&=&3\biggl[1+\sum_{n=2,4,\ldots} c_n^{(i)}\biggr],
\nonumber
\end{eqnarray}
and assuming that the numerical values of the Gegenbauer moments for the singlet and octet DAs are the same, 
we arrive at the estimate in Eq.~(\ref{eq:SoftEst}). 

%
%%%%%%%%%%%%%%%%%%%%%%%%%%%%%%%%%%%%%%
\subsection{Twist-two contribution}
%%%%%%%%%%%%%%%%%%%%%%%%%%%%%%%%%%%%%
%
For our purposes it is convenient to write  the required imaginary part of
$F^{\rm QCD}_{\gamma^*\gamma^*\to M}(Q^2,q^2)$  as sum of terms corresponding to
the expansion of the meson DAs $\phi_{M}(x,\mu)$ in Gegenbauer polynomials. The twist-2 quark components of the spectral densities with 
NLO accuracy can be obtained from relevant expressions presented in our work \cite{Agaev:2010aq}. 
Thus we write, for the flavor-octet contribution,
\begin{eqnarray}
\label{partialGeg}
 \lefteqn{\frac{1}{\pi}\mathrm{Im} F^{\rm QCD(8)}_{\gamma^*\gamma^*\to M}(Q^2,-s) =}
\\
& = & \frac{f_{M}^{(8)}}{3\sqrt{6}}
\sum_{n=0}^\infty c_{n,M}^{(8)}(\mu)\left[\rho_n^{(0)}(Q^2,s) + \frac{C_F\alpha_s}
{2\pi}\rho_n^{(1)}(Q^2,s;\mu)\right].
\nonumber
\end{eqnarray}
The LO partial spectral density is proportional to the meson DA
\begin{equation}
\rho^{(0)}_{n}(Q^2,s)=\frac{2\varphi_{n}(x)}{Q^2+s},\qquad \varphi_{n}(x)=6x{\bar x}C_{n}^{3/2}(2x-1),
\label{def:phin}
\end{equation}
where $x =Q^2/(Q^2+s)$.

The NLO spectral density can be written in the following form:
\begin{eqnarray}
\rho _{n}^{(1)}(Q^{2},s;\mu) &=& 
\frac{1}{Q^{2}+s}\biggl\{ \biggr\{\!\! -3\Big[1+2\left( \psi (2)-\psi (2+n)\right) \Big]
  \nonumber \\
&&{} +\frac{\pi ^{2}}{3}-\ln ^{2}\left( \frac{\bar{x}}{x}\right) -
\frac{{\gamma }_{n}^{(0)}}{C_F}\ln \left( \frac{s}{\mu ^{2}}\right) \biggr\}
\varphi _{n}(x) 
\nonumber \\
&&{}+\frac{{\gamma }_{n}^{(0)}}{C_F}\int_{0}^{\overline{x}}du\frac{\varphi
_{n}(u)-\varphi _{n}(\bar{x})}{u-\overline{x}} 
\nonumber \\
&&{}-\biggl[ \int_{x}^{1}du\frac{\varphi _{n}(u)-\varphi _{n}(x)}{u-x}%
\ln \left( 1-\frac{x}{u}\right) 
\nonumber\\
&&{}+(x\to \bar{x})\biggr] \biggr\},
\label{eq:ABOP}
\end{eqnarray}
where $\gamma^{(0)}_n$ is the flavor-nonsinglet LO anomalous dimension (\ref{eq:anomdim0}).
%\begin{equation}
% \gamma^{(0)}_n \equiv  C_F \tilde \gamma^{(0)}_n\,.
%\end{equation}

The flavor-singlet quark contribution can be written similarly as
\begin{eqnarray}
 \lefteqn{\frac{1}{\pi}\mathrm{Im} F^{\rm QCD(1,q)}_{\gamma^*\gamma^*\to M}(Q^2,-s) =}
\\
& \!=\! & \frac{2f_{M}^{(1)}}{3\sqrt{3}}
\sum_{n=0}^\infty c_{n,M}^{(1)}(\mu)\left[\rho_n^{(0)}(Q^2,s) + \frac{C_F\alpha_s}{2\pi}\rho_n^{(1)}(Q^2,s;\mu)\right]
\nonumber
\end{eqnarray}
with the same functions $\rho_n^{(0)}(Q^2,s)$ and $\rho_n^{(1)}(Q^2,s;\mu)$,  
the difference being encoded in the decay constants $f_{M}^{(i)}$, the expansion coefficients $c_{n,M}^{(i)}$ and numerical factors.

In order to find the contribution of the gluon DA one has to calculate the 
relevant Feynman diagrams (Fig.~1) for light quarks in the loop and two non-zero photon virtualities, $Q^2$ and $q^2$.
One obtains, omitting the factor $C_F\alpha_s/4\pi$,
\begin{widetext}
\begin{eqnarray}
T_{H}^{g}\Big|_{\mathrm{light}}\left( u,Q^{2},q^{2}\right)  &=&
-\frac{1}{u^{2}\bar{u}^{2}(Q^{2}-q^{2})^{2}}\biggl\{ Q^{2}u^{2}  
%\nonumber\\&&\times 
\ln \left( \frac{uQ^{2}+\bar{u}q^{2}}{Q^{2}}\right) \left[ \ln \left(
\frac{uQ^{2}+\bar{u}q^{2}}{\mu ^{2}}\right) +\ln \left( \frac{Q^{2}}{\mu ^{2}}\right) \right]  
\nonumber\\
&&-q^{2}\bar{u}^{2}\ln \left( \frac{uQ^{2}+\bar{u}q^{2}}{q^{2}}\right) \left[ \ln \left( \frac{uQ^{2}+\bar{u}q^{2}}{\mu ^{2}}\right)
+\ln \left( \frac{q^{2}}{\mu ^{2}}\right) \right]  
\nonumber\\
&&+2\left[ Q^{2}u\left( 3\bar{u}-2\right) \ln \left( \frac{u Q^{2}+\bar{u}q^{2}}{Q^{2}}\right) 
  +q^{2}\bar{u}\left(2-3u\right) \ln \left( \frac{ uQ^{2}+\bar{u}q^{2}}{q^{2}}\right) \right] \biggr\}.
\label{twovirt}
\end{eqnarray}
It is not difficult to verify that the result in (\ref{twovirt})
reproduces the known expression~(\ref{lightbox}) in the limit $q^2\to 0$. 
The corresponding contribution to the spectral density reads, replacing $q^2 \to -s$,  
\begin{eqnarray}
\frac{1}{\pi}\mathrm{Im}T_{H}^{g}\Big|_{\mathrm{light}}\! ( u,Q^{2},-s) & = & 
-\frac{2x}{Q^2}\biggl\{ \frac{1}{u^2\bar{u}^2}\left[\Theta (u-x)
\left[(x\bar{u}^2+\bar{x}u^2)\ln\left(1-\frac{\bar{u}}{\bar{x}}\right)+u\bar{u}\right]\right]
\nonumber \\
&&{}+\left[\Theta(u-x)\frac{x}{u^2}-\Theta(x-u)\frac{\bar{x}}{\bar{u}^2}\right]\left[\ln \frac{Q^2}{\mu^2}+\ln\frac{\bar{x}}{x}-2\right] \biggr\} .
\label{imbox}
\end{eqnarray}
\end{widetext}
A recalculation of the heavy $c$-quark contribution is not needed since the corresponding 
spectral density is not affected by the LCSR modification.
Thus the result in Eq.~(\ref{eq:charm}) obtained for $q^2=0$ can be used as it stands.
   
The contributions of different Gegenbauer polynomials in the expansion of the two-gluon DA
\begin{align}
 \omega _{n}(u) & = 30u^{2}\overline{u}^{2}C_{n-1}^{5/2}(2u-1)
\end{align}
defined as
\begin{eqnarray}
\hspace*{-0.6cm}
 \rho _{n}^{g}(Q^{2},s;\mu ) &\!=\!& \frac{1}{\pi }\!\int_{0}^{1}\!\!du\,\mathrm{Im}
T_{H}^{g}\Big|_{\mathrm{light}}\!\!\left( u,Q^{2},-s\right)\, \omega _{n}(u), 
\end{eqnarray}
can readily be computed from the above expressions. We obtain for $n=2$ and $n=4$:
\begin{eqnarray}
\rho _{2}^{g}(Q^{2},s,\mu) &\!=\!& \frac{5 x}{Q^2}\biggl[
 - \frac{ {}^{gq}\!\gamma_2^{(0)}}{C_F}\left(\ln\frac{\bar x Q^2}{x \mu^2}-2\right)\varphi_2(x)
\nonumber\\
&&{} + \frac{5}{6} \bar{x}^2\big(65x^2-30 x+1\big)  
\biggr],
\nonumber\\
\rho _{4}^{g}(Q^{2},s,\mu) &\!=\! & \frac{5x}{Q^2}\biggl[
 -\frac{{}^{gq}\!\gamma_4^{(0)}}{C_F}\left(\ln\frac{\bar x Q^2}{x \mu^2}-2\right)\varphi_4(x)
\nonumber\\&&{}
+ \frac{14}{15}\bar{x}^2\big(1827 x^4-2457 x^3 +959x^2
\nonumber\\
&&{}-105x+1\big)\biggr],
\end{eqnarray}
where $\varphi_n(x)$ are defined in (\ref{def:phin}) and the respective quark-gluon mixing anomalous 
dimension appear, because the coefficient of $\ln Q^2/\mu^2$ in (\ref{imbox}) is just the evolution kernel $V^{qg}(x,u)$.

Collecting all factors, the final expression for the contribution of the light quark box diagrams to the spectral density 
takes the following form: 
\begin{eqnarray}
 \lefteqn{\frac{1}{\pi }\mathrm{Im}F_{\gamma ^{\ast }\gamma ^{\ast}\to M}^{\rm QCD(g)}(Q^{2},-s) = }
\nonumber\\
& = &\frac{2f_{M}^{(1)}}{3\sqrt{3}}\sum_{n=2}^{\infty }c_{n,M}^{(g)}(\mu )%
\frac{C_{F}\alpha _{s}}{2\pi }\rho _{n}^{g}(Q^{2},s;\mu )\,.
\end{eqnarray}
As mentioned above, the contribution of charm quarks does not need to be written in this form as it is not
affected by the LCSR subtraction. 

%%%%%%%%%%%%%%%%%%%%%%%%%%%%%%%%%%%%%%
\subsection{Higher twist and meson mass corrections}
%%%%%%%%%%%%%%%%%%%%%%%%%%%%%%%%%%%%% 

The bulk of the higher-twist corrections corresponding to the contributions of two-particle and three-particle 
twist-four DAs can be taken into account using the expressions given in Ref.~ \cite{Agaev:2010aq} with the 
substitution of pion DAs by their $\eta, \eta'$ counterparts. The latter have been studied previously 
in~\cite{Ball:1998je,Ball:2006wn} but, as we found, the results given there are not complete. 
The corresponding update is presented in Appendix~\ref{App:HT}. We take into account quark mass corrections in the relations between 
different matrix elements imposed by QCD equations of motion (EOM) and also consider, for the first time,    
anomalous contributions to the flavor-singlet twist-four DAs.

In addition, one has to take into account the contribution of the twist-three DA, which appears  
due to the nonvanishing strange quark mass, and an extra meson mass correction $\sim m_M^2$
coming from the expansion of the leading order amplitude. 

In the expressions given below we collect the results for the spectral densities for the higher-twist contributions
defined as
\begin{align}
\rho^{(i)}_M=\frac{1}{\pi}\,\mbox{Im}\,F_{\gamma^*\gamma^*\to M}^{QCD(i)}(Q^2,-s)\,.
\end{align}
The superscript $i=m,\,3,\,4$ corresponds to the meson mass, twist-three DA and twist-four DA contributions, respectively. 
All higher-twist contributions can most conveniently be written as sum of contributions of different quark flavors
\begin{align}
 \rho_M^{(i)}(Q^2,s) &=
 2 e_s^2 ~\rho^{(i),s}_M\!\left(Q^2,s\right) + \sqrt{2} \left( e_u^2 \!+\! e_d^2 \right)  \rho^{(i),q}_M\!\left(Q^2,s\right).
 \end{align}
The rewriting in terms of the parameters in the FKS-scheme is then done 
using Eqs.~(\ref{eq:QFt2}), (\ref{QFS:Tw3}) for the leading twist and the same 
transformation rules for the higher-twist matrix elements 
$f_{3M}^{(a)}$ and $f_M^{(a)} \delta_M^{2(a)}$ where $a=q,s$.

The meson mass correction to the contribution of the $n$-th Gegenbauer term in the expansion of the
leading-twist DA, cf.~(\ref{partialGeg}), takes the form  
\begin{eqnarray}
 \rho_{M,n}^{(m),a}(Q^2,s)=\frac{x^2}{Q^4}h^a_M\left(\xi_x\varphi_n(x)-x\bar{x}\frac{d}{dx}\varphi_n(x)\right).
\end{eqnarray}
Here we used a shorthand notation
$$\xi_x=2x-1$$
and made a substitution $m_M^2 f_M^{(a)}\to h_M^{(a)}$ motivated in Appendix~\ref{App:HT}, Eq.~(\ref{ansatz}),
for consistency with the calculation of twist-four contributions.

The contribution of the twist-three DA to NLO accuracy in the conformal expansion reads  
\begin{align}
 \rho^{(3),a}_M(Q^2,s) &=
 - \frac{x^2}{Q^4}\left( h^a_M \xi_x +60 m_a f^a_{3M}C_3^{1/2}(2x-1) \right) \,,
\end{align}
and the twist-four contribution, to the same accuracy, can be brought into the form 
\begin{eqnarray}
 \rho^{(4),a}_M(Q^2,s)&=& -\frac{x^2 \xi_x}{Q^4}\biggl\{ \frac{160}{3}f^a_M(\delta^a_M)^2 x\bar{x} 
\nonumber\\&&{}
+  m_af^a_{3M}\Big[60 - 210x\bar{x}\left(3-x\bar{x}\right) \Big]
\nonumber\\
 &&{}+h^a_M\biggl[1-x\bar{x}\Big(\frac{13}{6}-\frac{21}{2}x\bar{x}\Big) 
\nonumber\\&&{}
+ c_{2M}^{(a)}x\bar{x}\left(21-135 x\bar{x}\right)\biggr] \biggr\}. 
\end{eqnarray}
In all expressions $a=q,s$ and $x = Q^2/(s+Q^2)$.

%%%%%%%%%%%%%%%%%%%%%%%%%%%%%%%%%%%%%%
%\subsection{Twist-six contribution}
%%%%%%%%%%%%%%%%%%%%%%%%%%%%%%%%%%%%%

The twist-six contributions to the $\gamma^*\gamma\to \pi^0$ transition FF have been
calculated in the factorization approximation in Ref.~\cite{Agaev:2010aq}.  
The extension of these results to $\gamma^*\gamma\to \eta,\eta' $ is not immediate as in order
to include $SU(3)$ flavor violation effects we would have to recalculate all the diagrams keeping terms linear in the quark masses. 
These would lead in the factorization approximation to contributions proportional to the twist 2 distribution amplitude 
times quark condensate. We postpone this calculation to a forthcoming publication and prefer to neglect 
the twist-six contributions altogether since at this level we would only be able to include them consistently for the octet 
but not the singlet. Neglecting them amounts to an additional uncertainty at the level of 2-3 percent and we will see that 
neither theoretical nor experimental precision are up to now sufficient to make these terms relevant.

%%%%%%%%%%%%%%%%%%%%%%%%%%%%%%%%%%%%%%%%%%%%%%%%%%%%%%%%%%%%%%%%%%%%%%%%%%%%%%%%%%%%%%%%%%%%%%%%%%%
%
\section{Numerical analysis}
\label{sec:NUMERICS}
%
%%%%%%%%%%%%%%%%%%%%%%%%%%%%%%%%%%%%%%%%%%%%%%%%%%%%%%%%%%%%%%%%%%%%%%%%%%%%%%%%%%%%%%%%%%%%%%%%%%%

%%%%%%%%%%%%%%%%%%%%%%%%%%%%%%
\subsection{Sum rule parameters}
%%%%%%%%%%%%%%%%%%%%%%%%%%%%%%

All numerical results in this work are obtained using the two-loop running QCD
coupling with $\Lambda_{\rm QCD}^{(4)} = 326$~MeV and $n_f=4$ active flavors.
Validity of the FKS mixing scheme for the DAs is assumed at the 
renormalization scale $\mu_0=1$~GeV, $\alpha_s(\mu_0) = 0.494$.
Unless stated otherwise, we use the set of FKS parameters specified in Eq.~(\ref{FKSvalues-old}).
All given values of nonperturbative parameters refer to the same scale $\mu_0=1$~GeV.

A natural factorization and renormalization scale $\mu$ in the calculation
of the meson transition FFs with two large photon virtualities is given by the
virtuality of the quark propagator $\mu^2 \sim \bar x Q^2 +  x q^2$. 
If $q^2\to 0$, in the LCSR framework the relevant factorization
scale becomes $\mu^2 \sim \bar x Q^2 + x M^2$ or $\mu^2 \sim \bar x Q^2 + x s_0$
if $M^2\gg s_0$, see e.g.~\cite{Braun:1999uj}.
Note that the restriction $s < s_0$ in the first integral in (\ref{eq:22}) translates to $\bar x < s_0/(s_0+Q^2)$
and hence the quark virtuality remains finite $\mu^2 \simeq 2 s_0$ as $Q^2\to \infty$, in agreement with the interpretation
of this term as the ``soft'' contribution.
Using the $x$-dependent factorization
scale is inconvenient so that we replace $x$ by the average
$\langle x \rangle$ which is varied within a certain range:
\begin{equation}
 \mu^2 = \langle \bar x \rangle\, Q^2 + \langle x \rangle\,s_0\,,
\qquad 1/4 < \langle x \rangle < 3/4\,.
\label{par:scale}
\end{equation}

The choice of the Borel parameter in LCSRs is discussed in~\cite{Ali:1993vd,Ball:1997rj}.
The difference to the classical SVZ sum rules is that the twist expansion in LCSRs goes in powers of $1/(x M^2)$
rather than $1/M^2$.
Hence one has to use somewhat larger values of $M^2$ compared to the QCD sum rules for two-point correlation functions
in order to ensure the same hierarchy of contributions. We choose as the ``working window''
\begin{equation}
    1 < M^2 < 2~\text{GeV}^2
\end{equation}
and $M^2=1.5$~GeV$^2$ as the default value in our calculations.

We use  the standard value $s_0=1.5$~GeV$^2$ for the continuum threshold, and the range
\begin{equation}
 1.3 < s_0 < 1.7~\text{GeV}^2
\end{equation}
in the error estimates.
We did not attempt to consider corrections due to the finite width
of the $\rho, \omega$ resonances. The estimates
in Ref.~\cite{Mikhailov:2009kf} suggest that such corrections
may result in an enhancement of the form factor by 2-4\% in
the small-to-medium $Q^2$ region where the resonance part dominates.
We believe that such uncertainties are effectively covered by our
(conservative) choice of the continuum threshold.

Finally, we use the values of the twist-three parameters
$h_q$ and $h_s$~\cite{Beneke:2002jn} specified in Eq.~(\ref{hvalues}), and 
also use $\delta^{2(q)}_M= \delta^{2(s)}_M = 0.2\pm 0.04$~GeV$^2$ 
\cite{Novikov:1983jt,Bakulev:2002uc}
(at the scale 1 GeV) for the normalization parameter for
twist-4 DAs (\ref{eq:delta2}).

%
%%%%%%%%%%%%%%%%%%%%%%%%%%%%%%
\subsection{Models of DAs and comparison with the data}
%%%%%%%%%%%%%%%%%%%%%%%%%%%%%%
%

%
\begin{figure*}[tb]
\begin{center}
\includegraphics[width=.88\textwidth,clip=true]{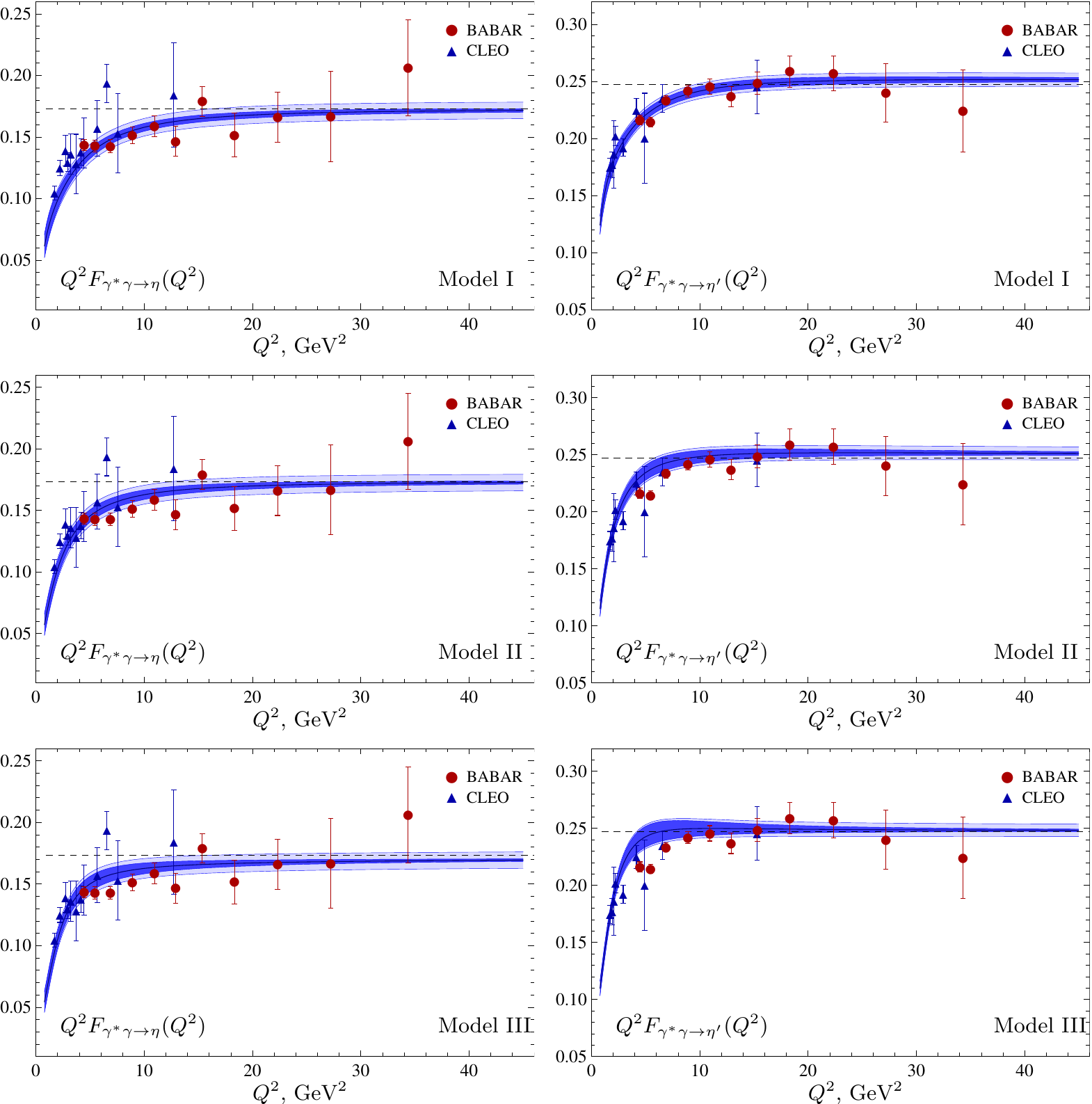}
\end{center}
\caption{Transition form factors $\gamma^*\gamma\to \eta$ (left panels) and 
$\gamma^*\gamma\to \eta'$ (right panels)~\cite{BABAR:2011ad,Gronberg:1997fj} compared to the LCSR calculation
with three models of the leading-twist DAs specified in Table~\ref{tab:models}.
Asymptotic values at large photon virtualities (\ref{eq:asy}) corresponding to the central values of the FKS parameters in 
Eq.~(\ref{FKSvalues-old}) are shown by the horizontal dashed lines. 
The dark blue shaded areas correspond to uncertainties of the calculation due to the choice of the LCSR parameters $M^2$ and $s_0$, 
factorization scale $\mu$ and the higher-twist parameters $h^{(q,s)}, \delta^{2(q,s)}$, see text.
The light blue areas are obtained by adding the uncertainties in the FKS parameters, Eq.~(\ref{FKSvalues-old}). 
}
\label{fig:LCSR-1}
\end{figure*}

The LCSR calculation of the FFs is compared with the experimental data~\cite{BABAR:2011ad,Gronberg:1997fj}
in Fig.~\ref{fig:LCSR-1}.
The dependence of the results on the Borel parameter, continuum threshold,
normalization of the higher-twist contributions and, to a lesser extent, 
the factorization scale, can be viewed as an intrinsic irreducible uncertainty 
of the LCSR method. This uncertainty is shown in the figures by the dark blue bands.

In this work we use the FKS mixing scheme~\cite{Feldmann:1998vh} as the simplest working hypothesis
that allows one to reduce the number of parameters, assuming that it holds for complete wave functions,
alias also for the DAs, at an \emph{ad hoc} low scale $\mu_0 =1$~GeV. The error bands corresponding to 
adding the uncertainties of the FKS parameters as given in Eq.~(\ref{FKSvalues-old}) to the LCSR 
uncertainties specified above is shown by light blue bands. We assume that all errors are
statistically independent and add them in quadrature. 
We expect that the bulk of these uncertainties will be eliminated in future by using first-principle
lattice calculations of the couplings $f_\eta$, $f_{\eta'}$ that are not bound to any mixing scheme. 

Asymptotic values of the form factors for large photon virtuality for the central values of the FKS 
parameters in Eq.~(\ref{FKSvalues-old}) are shown by the horizontal dashed lines, cf.~Eq.~(\ref{eq:asy}).
The asymptotic value for $\gamma^*\gamma\to \eta'$  differs considerably from the one 
assumed in~\cite{BABAR:2011ad,Kroll:2002nt,Kroll:2013iwa}, which is an effect of the 
finite renormalization correction to the flavor-singlet contribution, see Eq.~(\ref{eq:finrem}).  
Note that experimental measurements for both $\eta$ and $\eta'$ FFs at large virtualities are consistent 
with the expected asymptotic behavior.

\begin{figure*}[tb]
\begin{center}
\includegraphics[width=.88\textwidth,clip=true]{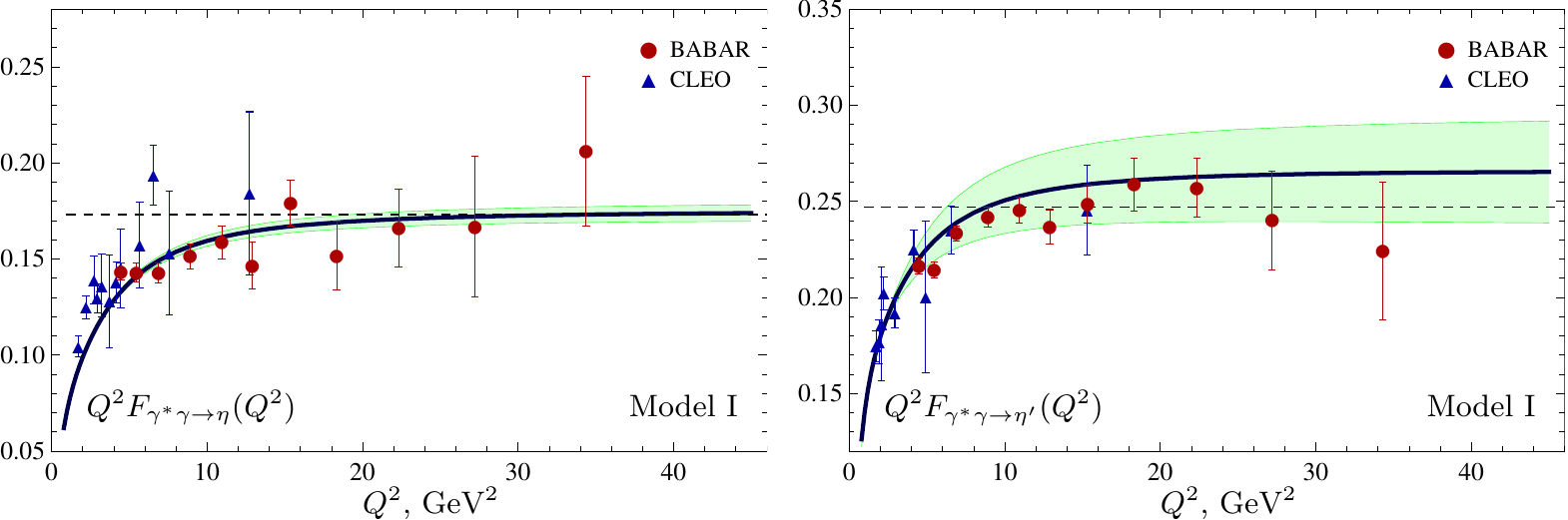}
\end{center}
\caption{Same as in Fig.~\ref{fig:LCSR-1} for the first model of the leading-twist DAs specified in Table~\ref{tab:models}
except for the normalization parameter of the gluon DA $c_2^{(g)}$ which is set to zero.
The shaded area in light green 
shows the effect of the variation of this parameter in the range $c_2^{(g)} = \pm 0.5$. 
}
\label{fig:LCSR-2}
\end{figure*}

The remaining nonperturbative input in the calculations is provided  
by the shape parameters of the DAs. We do not view this dependence as ``uncertainty''. 
Indeed, on the one hand, extraction of the information about DAs is the primary motivation behind the 
studies of transition form factors. On the other hand, lowest nontrivial moments of DAs can 
also be studied in lattice QCD \cite{Braun:2006dg,Arthur:2010xf}. Such calculations are ongoing
and the corresponding parameters will eventually be known to a sufficient precision.   

In the FKS approximation the remaining information about the DAs is encoded in three constants, 
$c_n^{(q)}(\mu_0)$, $c_n^{(s)}(\mu_0)$ and $c_n^{(g)}(\mu_0)$, for each Gegenbauer moment $n=2,4$, etc. 
The non-strange coefficients, $c_n^{(q)}(\mu_0)$, should be similar to the corresponding coefficients
for the pion DA. Unfortunately the situation with the pion DA is far from being settled. Direct calculations 
using QCD sum rules and lattice simulations do not have sufficient accuracy so far, whereas the 
constraints from the experimental data on the $\gamma^*\gamma\to \pi^0$ FF are inconclusive
because of the discrepancy between the BaBar and Belle measurements~\cite{Aubert:2009mc,Uehara:2012ag}. 
A detailed discussion can be found in~\cite{Agaev:2010aq,Agaev:2012tm}.  
    
Because of this uncertainty, we present the results for three different models
of the DAs specified in Table~\ref{tab:models} where the coefficients $c_n^{(q)}(\mu_0)$
are chosen in the range that correspond to popular models for the pion DA, 
the $SU(3)$-breaking in these parameters is neglected (see below),
and the gluon coefficients are fitted to describe the data. 
\begin{table}[ht]
\renewcommand{\arraystretch}{1.3}
\begin{center}
%\begin{tabular}{@{}l|l|l|l|l|l@{}} \hline
\begin{tabular}{|c|c|c|c|c|c|c|c|} \hline
  Model & $c^{(q)}_2$ & $c^{(s)}_2$ & $c^{(q)}_4$ & $c^{(s)}_4$ & $c^{(g)}_2$ 
\\ \hline
 I  & 0.10 & 0.10  & 0.10  & 0.10 & -0.26 
\\ \hline
 II & 0.20 & 0.20  & 0.0   & 0.0  & -0.31 
\\ \hline
III & 0.25 & 0.25 & -0.10 & -0.10 & -0.25
\\ \hline
\end{tabular}
\end{center}
\caption[]{Gegenbauer coefficients of three sample models of the leading-twist DAs
        {}[$u,d$-quarks (q), $s$-quarks (s) and gluons (g)] at the scale $\mu_0=1$~GeV. 
cf. Fig.~\ref{fig:LCSR-1}.} 
\label{tab:models}
\renewcommand{\arraystretch}{1.0}
\end{table}
The first model corresponds to the pion DA used in Ref.~\cite{Agaev:2012tm}
to describe the Belle data~\cite{Uehara:2012ag} (truncated to $n=2,4$), the second (simplest) model corresponds to a typical
ansatz used in vast literature on the weak $B\to\pi$ decays, and the third model with a negative $n=4$
coefficient is advocated by the Bochum-Dubna group, see e.g.~\cite{Bakulev:2012nh} and references therein.  

On general grounds one expects~\cite{Chernyak:1983ej} that the DAs of hadrons 
containing strange quarks are more narrow than those built of $u,d$ quarks, i.e. 
\begin{align}
  c_n^{(s)}(\mu_0) < c_n^{(q)}(\mu_0)\,, 
\end{align}  
however, existing numerical estimates of this effect are rather uncertain. 
QCD sum rule calculations (see e.g.~\cite{Ball:1998je,Ball:2006wn}) and 
lattice calculations~\cite{Braun:2006dg,Arthur:2010xf} do not seem to indicate any large
difference so that we have assumed $c_n^{(s)}(\mu_0) = c_n^{(q)}(\mu_0)$ for the present study.
Setting instead $c_n^{(s)}(\mu_0)=0$, which is probably extreme, the FF $\gamma^*\gamma\to \eta$ gets increased
by 5-6\% and the FF $\gamma^*\gamma\to \eta'$ decreases by 4-5\% for $Q^2> 5$ GeV$^2$ 
as compared to the results shown in Fig.~\ref{fig:LCSR-1}.    
  
The gluon DA mainly contributes to the $\eta'$ FF, whereas its effect on the $\eta$ is small. 
To illustrate this dependence we show in Fig~\ref{fig:LCSR-2} the results of the calculation 
with $c^{(q)}_2 =  c^{(q)}_4 = 0.1$ and $c^{(g)}_2 = 0$ corresponding to Model I with gluon contribution 
put to zero (blue curve), and the the shaded area in light green obtained by varying $c^{(g)}_2$
in the range $-0.5 < c^{(g)}_2  < 0.5$. Note that the gluon DA contribution is significantly enhanced 
(by a factor 5/3 for large $Q^2$) by including the $c$-quark contribution, which is one of the new 
elements of our analysis.

\begin{figure*}[tb]
\begin{center}
\includegraphics[width=.88\textwidth,clip=true]{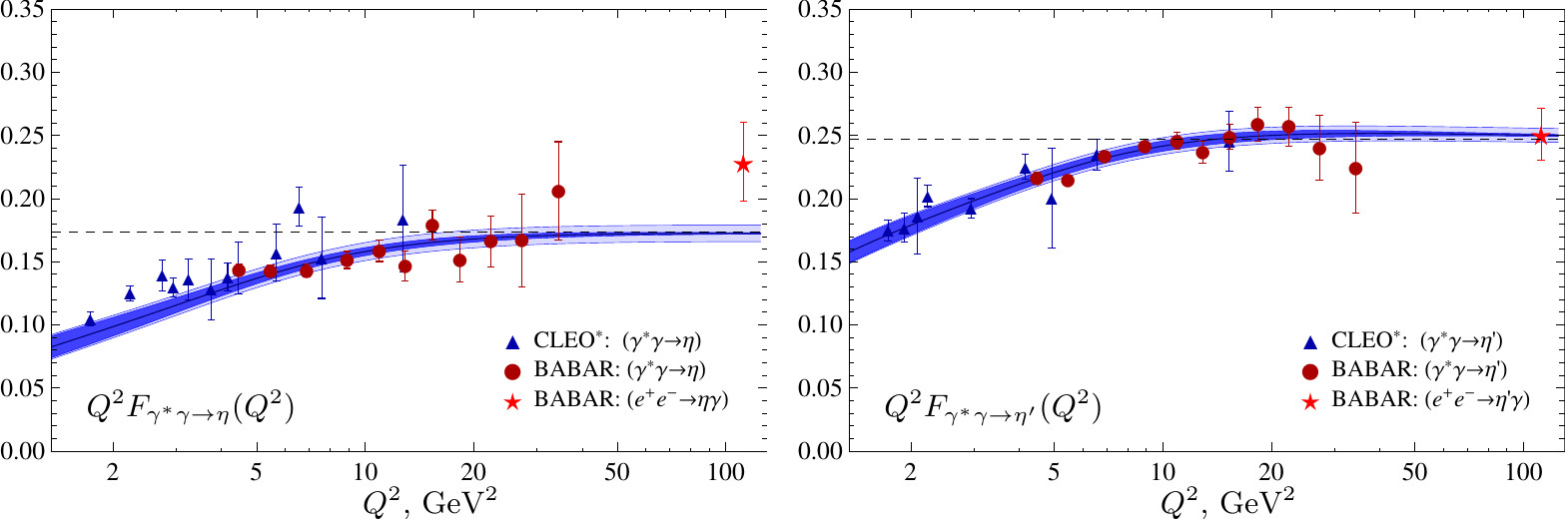}
\end{center}
\caption{Same as in Fig.~\ref{fig:LCSR-1} using a logarithmic scale in $Q^2$.  
The calculation uses the first model of the leading-twist DAs specified in Table~\ref{tab:models}.
The time-like data point~\cite{Aubert:2006cy} at $|Q^2|= 112\,\,\mathrm{GeV}^{2}$ is shown by red stars
for comparison.
}
\label{fig:LCSR-3}
\end{figure*}

The three models in Table~\ref{tab:models} lead to an equally good description of the 
experimental data at large $Q^2>10-15$~GeV$^2$ but differ at smaller $Q^2$ where Model I 
seems to be preferred. Unfortunately, the uncertainties of the calculation also increase in this region, 
especially for Model III which suffers from a stronger dependence on the Borel parameter.     
For this reason we think that none of the considered models can be excluded and, also in future, 
the experimental data on transition FFs alone will not be sufficient to pin down the shape of DAs.
One needs a combined effort of theory and experiment, 
supplementing FF data with lattice calculations of at least a few   
key parameters.

Finally, in Fig.~\ref{fig:LCSR-3} we show the same results on a logarithmic scale in $Q^2$, where we have
also included the time-like momentum transfer data point~\cite{Aubert:2006cy} at $|Q^2|= 112\,\,\mathrm{GeV}^{2}$ (red stars)
for comparison.    

One sees that the measurement of $e^+ e^-\to\gamma^*\to \eta'\gamma$ appears to be in good agreement with the 
expected asymptotic behavior in the space-like region, whereas 
the result for $e^+ e^-\to\gamma^*\to \eta'\gamma$ is considerably higher.
This difference is interesting and surprising. The Sudakov enhancement of the time-like FFs as compared to their space-like
conterparts,  usually quoted in this context, is universal and 
should affect both $\eta$ and $\eta'$ production equally strongly. As already discussed in Section~III.C,  
the large difference can only be attributed to nonperturbative corrections corresponding 
to the soft (end-point) integration regions. Although a rigorous connection of such contributions to the DAs
does not exist, one can plausibly argue that large soft corrections are correlated with the end-point
enhancements in the DAs, of the type that have been discussed in connection with the large scaling violation 
in the $\gamma^*\gamma\to\pi^0$ form factor reported in~\cite{Aubert:2009mc}. 
{}For this reason we expect that, if the large value of the time-like form factor for the $\eta$ meson
is confirmed, the corresponding space-like form factor should exibit the similar scaling violating
behavior as observed by BaBar for the pion. In fact the existing data
may support such a trend, see Fig.~\ref{fig:LCSR-3}, although it is not statistically significant.

%%%%%%%%%%%%%%%%%%%%%%%%%%%%%%%%%%%%%%%%%%%%%%%%%%%%%%%%%%%%%%%%%%%%%%%%%%%%%%%%%%%%%%%%%%%%%%%%%%%
%
\section{Summary and Conclusions}
\label{sec:SUMMARY} 
%
%%%%%%%%%%%%%%%%%%%%%%%%%%%%%%%%%%%%%%%%%%%%%%%%%%%%%%%%%%%%%%%%%%%%%%%%%%%%%%%%%%%%%%%%%%%%%%%%%%%

In anticipation for the possibility of high-precision measurements of the 
transition form factors $\gamma^*\gamma\to\eta$ and $\gamma^*\gamma\to\eta'$ at the upgraded 
KEKB facility, in this work we update the corresponding theoretical framework.
The presented formalism incorporates several new elements in comparison to the existing 
calculations, in partucular a full NLO analysis of perturbative corrections, 
the charm quark contribution, and revisited twist-four contributions taking into account
$SU(3)$-flavor breaking and the axial anomaly.
A numerical analysis of the existing experimental data is performed with these improvements. 

For the numerical analysis we have used the FKS state mixing assumption for the $\eta$, $\eta'$ DAs 
at a low scale 1 GeV as a working hypothesis to avoid proliferation of parameters. 
This assumption does not contradict the data on the FFs at small-to-moderate photon virtualities
and can be relaxed in future, if necessary. 

The most important effect of the NLO improvement is due to the finite renormalization of 
the flavor-singlet axial current which results in a  20\% reduction of the the expected 
asymptotic value of the $\gamma^*\gamma\to\eta'$ form factor at large photon virtualities.
Taking into account this correction brings the result in agreement with BaBar 
measurements~\cite{BABAR:2011ad}.

We also want to emphasize the importance of taking into account the charm quark contribution.
This effect is negligible at small $Q^2$, but increases the contribution of the most interesting 
two-gluon DA by a factor 5/3 at large scales, so that a consistent implementation of the c-quark mass 
threshold effects is mandatory.

The update of the higher-twist corrections does not have a large numerical impact, but is
necessary for theoretical consistency with taking into account the meson mass corrections 
to the leading-twist diagrams. Identifying the hadron mass corrections in hard exclusive reactions 
is in general a nontrivial problem~\cite{Braun:2011dg}, and it is made even harder by the axial anomaly.    
We have calculated the anomalous contribution to the twist-four DA for one particular case
and found a specific mechanism how this contribution can restore the relations 
between $\eta$, $\eta'$ masses implied by the state-mixing assumption for higher-twist. 
  
Our results for the FFs at Euclidean virtualities 
%obtained using the DAs inspired by the pion FF data, see Table.~\ref{tab:models},
are, in general, in good agreement with
the experimental data~\cite{BABAR:2011ad}, although the present statistical accuracy
of the measurements is insufficient to distinguish between different models of the DAs
specified in Table.~\ref{tab:models}. 
We expect that experimental errors will become smaller in future, and also that some of the 
parameters, most importantly the decay constants $f_\eta$, $f_{\eta'}$, will be calculated 
with high precision on the lattice. In this way the comparison of the QCD calculation with 
experiment will allow one to study the structure of $\eta$, $\eta'$ mesons at short
interquark separations, encoded in the DAs, on a quantitative level. 
  
We have given a short discussion of the transition form factors in the time-like region $q^2 = -Q^2 > 0$. 
The result by BaBar~\cite{Aubert:2006cy} suggesting a large enhancement of the $\eta$ form factor 
in the time-like as compared to the space-like region, and at the same time no such enhancement 
for $\eta'$ is rather puzzling. 
If confirmed, this difference would imply a significant 
difference in the end-point behavior of $\eta$ and $\eta'$ DAs.

%%%%%%%%%%%%%%%%%%%%%%%%%%%%%%%%%%%%%%%%%%%%%%%%%%%%%%%%%%%%%%%%%%%%%%%%%%%%%%%%%%%%%%%%%%%%%%%%%%%
%
\section*{Acknowledgments} 
%
%%%%%%%%%%%%%%%%%%%%%%%%%%%%%%%%%%%%%%%%%%%%%%%%%%%%%%%%%%%%%%%%%%%%%%%%%%%%%%%%%%%%%%%%%%%%%%%%%%%

This project was supported by Forschungszentrum J\"ulich (FFE contract 42008319 (FAIR-014)) and DAAD (grant A/13/03701).
The work of S.S. Agaev was also supported by Grant EIF-Mob-3-2013-6(12)-14/01/1-M-02 of the Science Development Foundation of the President of the Azerbaijan Republic.

%%%%%%%%%%%%%%%%%%%%%%%%%%%%%%%%%%%%%%%%%%%%%%%%%%%%%%%%%%%%%%%%%%%%%%%%%%%%%
%%%%%%%%%%%%%%%%%%%%%%%%%%%%%   Appendix   %%%%%%%%%%%%%%%%%%%%%%%%%%%%%%%%%%
%%%%%%%%%%%%%%%%%%%%%%%%%%%%%%%%%%%%%%%%%%%%%%%%%%%%%%%%%%%%%%%%%%%%%%%%%%%%%

\appendix 

\renewcommand{\theequation}{\Alph{section}.\arabic{equation}}

%%%%%%%%%%%%%%%%%%%%%%%%%%%%%%%%%%%%%%%%%%%%%%%%%%%%%%%%%%%%%%%%%%%%%%%%%%%%%%%%%%%%%%%%%%
%
\section{DAs of twist four}
\label{App:HT} 
%
%%%%%%%%%%%%%%%%%%%%%%%%%%%%%%%%%%%%%%%%%%%%%%%%%%%%%%%%%%%%%%%%%%%%%%%%%%%%%%%%%%%%%%%%%%

This Appendix contains a detailed discussion and an update of the twist-four DAs of pseudoscalar mesons.  
To this end we follow the classification and notations in Ref.~\cite{Ball:2006wn} adapted for
our present case. The presentation is divided into two parts. In the first subsection we ignore 
anomalous contributions. This part contains the necessary definitions and an update of the results 
in~\cite{Ball:1998je,Ball:2006wn} taking into account quark mass corrections in the relations
between different matrix elements.   
The given expressions can be used as written for the flavor-octet contributions but have to be modified for 
flavor-singlet ones. Anomalous contributions to the flavor-singlet twist-four DAs are considered in the 
second subsection. This is an entirely new subject; we are not aware of any related studies beyond twist-two 
accuracy. The complete solution requires a full NLO evaluation of twist-four contributions and 
goes beyond the scope of this work. Instead, we formulate a simple substitution rule that is 
based on a sample calculation of the anomaly for one particularly important case, 
and is likely to take into account the bulk of the effect.    
  
%%%%%%%%%%%%%%%%%%%%%%%%%%%%%%%%%%%%%%%%%%%%%%%%%%%%%%%%%%%%%%%%%%%%%%%%%%%%%%%%%%%%%%%%%%
%
\subsection{General classification and quark mass corrections}
\label{App:HTa} 
%
%%%%%%%%%%%%%%%%%%%%%%%%%%%%%%%%%%%%%%%%%%%%%%%%%%%%%%%%%%%%%%%%%%%%%%%%%%%%%%%%%%%%%%%%%%

There exist four different three-particle twist-four DAs that can be defined as, e.g. for the strange quarks  
\begin{eqnarray}
\lefteqn{\hspace*{-0.5cm}\langle 0 | \bar s (z_2 n)\gamma_\mu\gamma_5
gG_{\alpha\beta}(z_3 n) s (z_1 n)|M(p)\rangle\ =}
\hspace*{0.5cm}\nonumber\\
& = & p_\mu (p_\alpha n_\beta - p_\beta n_\alpha)\, \frac{1}{pn}\, F^{(s)}_{M} \Phi^{(s)}_{4;M}(\underline{z},pn) 
\nonumber\\&&{}
+ (p_\beta g_{\alpha\mu}^\perp - p_\alpha g_{\beta\mu}^\perp) F^{(s)}_{M} \Psi^{(s)}_{4;M}(\underline{z},pn) + \dots,
\nonumber\\
\lefteqn{\hspace*{-0.5cm}\langle 0 | \bar s (z_2n)\gamma_\mu i
g\widetilde{G}_{\alpha\beta}(z_3n)s(z_1 n)| M(p)\rangle\ =}
\nonumber\\
& = & p_\mu (p_\alpha n_\beta - p_\beta n_\alpha)\, \frac{1}{pn}\, F^{(s)}_{M}
\widetilde\Phi^{(s)}_{4;M}(\underline{z},pn) 
\nonumber\\&&{}
+ (p_\beta g_{\alpha\mu}^\perp -
p_\alpha g_{\beta\mu}^\perp) F^{(s)}_{M} \widetilde\Psi^{(s)}_{4;M}(\underline{z},pn) + \dots,
\label{eq:DA-T4-3p}
\end{eqnarray}
with the short-hand notation
\begin{align}
{\cal F}(\underline{z},pn) &= \int{\cal D}\underline{\alpha}\,
e^{-ipz(\alpha_1 z_1 + \alpha_2 z_2+\alpha_3 z_3 )} 
{\cal F}(\underline{\alpha})\,,
\notag\\
\int{\cal D}\underline{\alpha} &= \int_0^1 d\alpha_1 d\alpha_2 d\alpha_3  \delta\big(1-\sum \alpha_i\big)
\end{align}
and $g^\perp_{\alpha\mu} = g_{\alpha\mu} - (p_\alpha n_\mu + p_\mu n_\alpha)/(pn)$, etc.  The ellipses stand for contributions
of twist higher than four. C-parity implies that the DAs $\Phi$ and $\Psi$ are 
antisymmetric under the interchange of the quark momenta, $\alpha_1\leftrightarrow \alpha_2$, 
whereas $\widetilde\Phi$ and $\widetilde\Psi$ are symmetric.
The three-particle twist-four DAs for $q=(u,d)$ quarks are defined by the same expressions with 
obvious substitution of the quark fields and the superscripts $(s)\to (q)$,
% and with an extra $1/\sqrt{2}$ factor in the overall normalization, 
cf.~Eqs.~(\ref{eq:DA-LT}).   

Three-particle DAs can be expanded in orthogonal polynomials that correspond to contributions of increasing spin
in the conformal expansion. Taking into account contributions of the lowest and the next-to-lowest spin 
one obtains~\cite{Braun:1989iv,Ball:1998je,Ball:2006wn}
\begin{eqnarray}
\Phi_{4;M}(\underline{\alpha}) & = & 120 \alpha_1\alpha_2\alpha_3 \Big[\phi_{1,M}^{(s)} (\alpha_1-\alpha_2)\Big],
\nonumber\\
\widetilde\Phi_{4;M}(\underline{\alpha}) & = & 120
\alpha_1\alpha_2\alpha_3 \Big[ \widetilde\phi_{0,M}^{(s)} +
\widetilde\phi_{2,M}^{(s)} (3\alpha_3-1)\Big],
\nonumber\\
{\widetilde\Psi}^{(s)}_{4;M}(\underline{\alpha}) & = &
 -30 \alpha_3^2\Big\{ \psi^{(s)}_{0,M}(1-\alpha_3)
\nonumber\\&&{}
                \hspace*{0.8cm}    
+\psi^{(s)}_{1,M}\Big[\alpha_3(1\!-\!\alpha_3)-6\alpha_1\alpha_2\Big]
\nonumber\\&&{}
                 \hspace*{0.8cm}   +\psi^{(s)}_{2,M}\Big[\alpha_3(1\!-\!\alpha_3)-\frac{3}{2}(\alpha_1^2
                               +\alpha_2^2)\Big]\Big\},
\nonumber\\
 {\Psi}^{(s)}_{4;M}(\underline{\alpha}) & = &
 - 30 \alpha_3^2 (\alpha_1-\alpha_2) \Big\{ 
                    		    \psi^{(s)}_{0,M} + \psi^{(s)}_{1,M} \alpha_3  
\nonumber\\&&{}  \hspace*{2.5cm}
+ \frac{1}{2} \psi^{(s)}_{2,M}(5 \alpha_3-3)\Big\}.
\label{eq:T4-conformal}
\end{eqnarray} 
The coefficients $\phi_{k,M}^{(s)}$, $\psi_{k,M}^{(s)}$ are related by QCD equations of motion (EOM)~\cite{Braun:1989iv}. One such relation is rather 
nontrivial and involves the divergence (in the mathematical sense) of the spin-three conformal operator
\begin{eqnarray}
\mathbb{O}^{(\bar s s)}_{\mu\alpha\beta} &=&
 \bar s \der_\alpha\der_\beta \gamma_\mu\gamma_5 s 
-\frac15 \partial_\alpha\partial_\beta \bar s \gamma_\mu \gamma_5 s\,,
\end{eqnarray}
where the symmetrization in all Lorentz indices and subtraction of traces are implied. 
Ignoring possible anomalous contributions to be discussed later, we obtain
\begin{eqnarray}
6\,\partial^\mu \mathbb{O}^{(\bar s s)}_{\mu\alpha\beta} &=&
-24i \bar s \gamma^\rho\Big(G_{\rho\beta}\derright_\alpha -\derleft_\alpha G_{\rho\beta}\Big)\gamma_5 s
\nonumber\\&&{}
+ 4i m_s \bar s \der_\alpha\der_\beta \gamma_5 s
-  16 i m_s \bar s \sigma^{\alpha\rho}G_{\rho\beta}\gamma_5s
\nonumber\\&&{}
-\frac{16}{3}  \partial_\beta \bar s \gamma^\rho \widetilde G_{\rho\alpha}s 
-8  \partial^\rho \bar s \gamma_\beta \widetilde G_{\alpha\rho} s 
\nonumber\\&&{}
-\frac{4}{15} im_s \partial_\alpha\partial_\beta \bar s \gamma_5 s
-\,\text{traces}
%\nonumber\\&&{}
% - \frac43 i(m_d-m_u) \partial_\beta \bar u \der_\alpha\gamma_5 d    
%-\,\text{traces}
\label{eq:divergence}
\end{eqnarray}
The quark-mass corrections in this expression $\sim \mathcal{O}(m_s)$ are a new result; 
they have not been taken into account in~\cite{Ball:1998je,Ball:2006wn}.

After some algebra we obtain
\begin{align}
  \widetilde\phi_{0,M}^{(s)} = \psi^{(s)}_{0,M} = - \frac13\, \delta^{2(s)}_{M}\,,   
\end{align}
where the parameter $\delta^{2(s)}_{M}$ is defined as 
\begin{align}
 \langle 0| \bar s \gamma^\rho ig \widetilde{G}_{\rho\mu} s |M(p)\rangle = p_\mu f_M^{(s)} \delta^{2(s)}_{M}\,,    
\label{eq:delta2}
\end{align}
and
\begin{align}
  \widetilde\phi_{2M}^{(s)}   &= \frac{21}{8} \delta^{2(s)}_{M}  \omega_{4M}^{(s)}, 
\notag\\
  \phi_{1M}^{(s)}    &=  \frac{21}{8}\left[\delta^{2(s)}_{M}\omega_{4M}^{(s)}   + \frac{2}{45} m^2_M \left(1-\frac{18}{7}c^{(s)}_{2M}\right) \right],
\notag\\
%  \psi_1^\pi + \psi_2^\pi &=  \frac{21}{4} \delta_M^2  \omega_{4M}^{(s)} 
%\notag\\
%  \psi_1^\pi - \psi_2^\pi &=
%   \frac{7}{4} \left[
% - \delta_M^2\omega_{4M}^{(s)} + \frac{2}{45} m^2_M \left(1\!-\!\frac{18}{7}a^{(s)}_2\right) + 4 (m_u+m_d) \frac{f_{3M}^{(s)}}{f_M}  
%                \right]
 \psi_{1M}^{(s)} & =    \frac{7}{4} \left[
  \delta^{2(s)}_{M}\omega_{4M}^{(s)} \!+\! \frac{1}{45} m^2_M \left(1\!-\!\frac{18}{7}c^{(s)}_{2M}\right) \!+\! 4m_s \frac{f_{3M}^{(s)}}{f^{(s)}_M}  
                \right]\!,  
\notag\\
 \psi_{2M}^{(s)} & =    \frac{7}{4} \left[
 2 \delta^{2(s)}_{M}\omega_{4M}^{(s)} \!-\! \frac{1}{45} m^2_M \left(1\!-\!\frac{18}{7}c^{(s)}_{2M}\right) \!-\! 4 m_s \frac{f_{3M}^{(s)}}{f^{(s)}_M}  
                \right]\!, 
\label{NLOspin}
\end{align}
where
\begin{eqnarray}
\lefteqn{\langle 0 | \bar s [iD_\mu,ig\widetilde{G}_{\nu\xi}] \gamma_\xi s -
\frac{4}{9}\, i\partial_\mu \bar s i g
\widetilde{G}_{\nu\xi}\gamma_\xi s | M (p)\rangle\
=}\hspace*{0.5cm}\nonumber\\
& = & f^{(s)}_M\delta^{2(s)}_{M} \omega_{4M}^{(s)} \left(p_\mu p_\nu - \frac{1}{4}\, m_M^2
g_{\mu\nu}\right) + {\mathcal O}({\rm twist\ 5}).
\nonumber\\
\end{eqnarray}
The expressions in (\ref{NLOspin}) differ from those in~\cite{Ball:1998je,Ball:2006wn} in terms $\sim m^2_M$ that arise from the quark mass corrections
in the divergence of the conformal operator~(\ref{eq:divergence}) and, surprisingly, also in terms $\sim m^2_M c^{(s)}_{2M}$:
The result for such terms obtained in~\cite{Ball:1998je} (and used in~\cite{Ball:2006wn}) is 
recovered if in our expressions $m^2_M c^{(s)}_{2M} \to (3/2)  m^2_M c^{(s)}_{2M}$. 

In addition one defines the two-particle twist-4 DAs as corrections $\sim\mathcal{O}(x^2)$ in the 
light-cone expansions $x^2\to 0$ of the nonlocal matrix element
\begin{eqnarray}
\lefteqn{
\langle 0 | \bar{s}(z_2 x) \gamma_\mu\gamma_5 s(z_1 x) |M(p)\rangle =}
\nonumber\\ &=& i\,p_\mu F_{M}^{(s)}\int_{0}^{1}\!\!du\, e^{-iz_{21}^{u}(px)}
\Big[\phi_{M}^{(s)}(u)
%\nonumber\\&&\hspace*{3cm}{}
+\frac{z_{12}^2 x^2}{16} \phi^{(s)}_{4M}(u)\Big] 
\nonumber \\
&&\quad + \frac{i}{2} \frac{x_\mu}{(px)} F_{M}^{(s)} \int_{0}^{1}\!\!du\, e^{-iz_{21}^{u}(px)} \psi_{4M}^{(s)}(u)\,. 
\label{eq:DA-T4-2p}
\end{eqnarray}
The DAs $\phi^{(s)}_{4M}(u)$, $\psi^{(s)}_{4M}(u)$ can be calculated in terms of the three-particle 
DAs of twist four and the DAs of lower twist defined in the main text, 
making use of the operator identities (see e.g.~\cite{Ball:2006wn}) 
\begin{eqnarray}
\label{eq:oprel1}
\lefteqn{
\frac{\partial}{\partial x^\mu}\, \bar s(x)[x,-x]\gamma_\mu\gamma_5 s(-x) =}
\\
& = & - i\! \int_{-1}^1\!\! dv\, v\, \bar s (x)[x,vx] x^\alpha
gG_{\alpha\mu}(vx) \gamma^\mu\gamma_5 [vx,-x]s(-x)\,,
\nonumber
\end{eqnarray}
and
\begin{eqnarray}
\lefteqn{\partial_\mu \{\bar s(x)[x,-x]\gamma^\mu\gamma_5 s(-x)\} =}
\nonumber\\
& =& - i\int_{-1}^1\!\! dv\, \bar s(x)[x,vx] x^\alpha
gG_{\alpha\mu}(vx) \gamma^\mu\gamma_5[vx,-x] s(-x)
\nonumber\\
&& {} + 2m_s\bar s(x)[x,-x]i\gamma_5 s(-x),
\label{eq:oprel2}
\end{eqnarray}
where $[x,y]$ is the straight-line-ordered Wilson line connecting 
the points $x,y$ and $\partial_\mu$ is the total derivative defined as
\begin{eqnarray}
\lefteqn{\partial_\mu \left\{ \bar u(x)\Gamma d(-x)\right\} \equiv }
\nonumber\\&\equiv&
\left.\frac{\partial}{\partial y_\mu}\,\left\{ \bar u(x+y) [x+y,-x+y]
    \Gamma d(-x+y)\right\}\right|_{y\to 0}.
\end{eqnarray}
Taking the matrix elements of these identities and putting $x^2\to 0$ afterwards, one obtains
the expressions for two-particle DAs $\psi_{4M}^{(s)}(u)$ and $\psi_{4M}^{(s)}(u)$ that can conveniently 
be separated in ``genuine'' twist-four contributions and meson mass corrections as
\begin{align}
    \psi_{4M}^{(s)}(u) = \psi_{4M}^{(s){\rm twist}}(u) +  m^2_M \psi_{4M}^{(s){\rm mass}}(u)    
\end{align} 
with
\begin{eqnarray}
 \psi_{4M}^{(s){\rm twist}}(u) &=& \frac{20}{3} \delta^{2(s)}_M C_2^{1/2}(2u-1) + 30 m_s \frac{f^{(s)}_{3M}}{f^{(s)}_{M}} 
\nonumber\\&&{}
\times \Big(\frac{1}{2}-10 u\bar u +35 u^2\bar u^2\Big)\,,
\nonumber\\
 \psi_{4M}^{(s){\rm mass}}(u) &=& \frac{17}{12} - 19 u\bar u + \frac{105}{2} u^2\bar u^2 
\nonumber\\&&{}
+  c^{(s)}_{2,M} \Big(\frac{3}{2} - 54  u\bar u  + 225 u^2\bar u^2 \Big)
\end{eqnarray}
and similarly
\begin{align}
    \phi_{4M}^{(s)}(u) = \phi_{4M}^{(s){\rm twist}}(u) +  m^2_M \phi_{4M}^{(s){\rm mass}}(u)\,,    
\end{align} 
where
\begin{eqnarray}
 \phi_{4M}^{(s){\rm twist}}(u) &=& \frac{200}{3}\delta^{2(s)}_M u^2 \bar u^2  
+ 21 \delta^{2(s)}_M \omega_{4M}^{(s)}\Big\{ u\bar u (2\!+\!13 u\bar u)
\nonumber\\&&{}
+ 2\big[ u^3(10-15 u+6 u^2)\ln u + (u\leftrightarrow \bar u)\big]\Big\}
\nonumber\\&&{}
+ 20 m_s \frac{f^{(s)}_{3M}}{f^{(s)}_{M}} u\bar u
\Big[12 -63u \bar u + 14u^2 \bar u^2\Big],
\nonumber\\
 \phi_{4M}^{(s){\rm mass}}(u) &=&  u\bar u \Big[\frac{88}{15} + \frac{39}{5} u\bar u + 14 u^2\bar u^2\Big]
\nonumber\\&&{}
-  c^{(s)}_{2,M} u\bar u \Big[
\frac{24}{5} - \frac{54}{5} u\bar u + 180 u^2\bar u^2\Big]
\nonumber\\&&
+ \Big(\frac{28}{15}-\frac{24}{5} c^{(s)}_{2,M} \Big)\Big[ u^3(10-15 u+6 u^2)\ln u
\nonumber\\&&{}
 + (u\leftrightarrow \bar u)\Big].
\end{eqnarray}
These results supersede the corresponding expressions in Ref.~\cite{Ball:2006wn,Khodjamirian:2009ys}.

%%%%%%%%%%%%%%%%%%%%%%%%%%%%%%%%%%%%%%%%%%%%%%%%%%%%%%%%%%%%%%%%%%%%%%%%%%%%%%%%%%%%%%%%%%
%
\subsection{Anomalous contributions}
\label{App:HTanomalous} 
%
%%%%%%%%%%%%%%%%%%%%%%%%%%%%%%%%%%%%%%%%%%%%%%%%%%%%%%%%%%%%%%%%%%%%%%%%%%%%%%%%%%%%%%%%%%

The general reason why the results in the previous subsection are incomplete is that the
operator identities (\ref{eq:divergence}), (\ref{eq:oprel1}), (\ref{eq:oprel2}) are valid
in this form only for bare (unrenormalized) operators. The renormalization $Z$-factor for the 
light-ray operator on the l.h.s. of, e.g., Eq.~(\ref{eq:oprel2}) can be written as an integral
operator acting on the field coordinates, see~\cite{Balitsky:1987bk}. The derivative $\partial_\mu$
can be brought inside the integral so that the algebra leading to the expression on the 
r.h.s. of this equation remains unchanged. However, the result is not yet written in terms of 
renormalized operators. Since the overall expression is finite (as a derivative of a finite 
operator) it can further be re-expanded in contributions of renormalized operators. 
In this way the coefficient functions of the operators that are already present will be modified
by $\alpha_s$ corrections and \emph{all other} operators with proper quantum numbers can appear, 
with coefficient functions starting at order $\mathcal{O}(\alpha_s)$.
Whereas this complication is, generally speaking, only relevant if the calculation of 
twist-four corrections is done to NLO accuracy (in which case the $\alpha_s$ corrections
to the coefficient functions of the OPE of the product of two electromagnetic currents
have to be taken into account as well), the contribution of gluon operators related to the
axial anomaly deserves special attention because of its role in the pattern of chiral symmetry
breaking for pseudoscalar mesons.    

To begin with, we recall the derivation of the celebrated anomaly relation (\ref{anomaly}) 
for the axial current:
\begin{align}
 \partial_\mu \bar s \gamma^\mu\gamma_5 s = 2m_s\bar s i\gamma_5 s  
   + \bar s \Big[ (\sderleft -i m_s)\gamma_5  - \gamma_5(\sderright + i m_s) \Big] s\,. 
\end{align}
The EOM terms (Dirac operator applied to a quark field) can be substituted inside the QCD path integral
by a functional derivative with respect to the corresponding antiquark field,
\begin{align}
     (\sderright + i m_s)s(y) e^{iS_\psi} =  - \frac{\delta}{\delta\bar s(y)} e^{iS_\psi},
\end{align}
where $S_\psi$ is the fermion part of the action. Such contributions can usually be dispensed of
by partial integration inside the path integral, producing contact terms. Anomalous 
contributions arise when the derivative $\delta/\delta\bar s(y)$ acts on the 
antiquark field in the same composite operator, in our case the axial current, producing
ill-defined contributions $\sim \delta^4(0)$ that have to be regularized.                  

A well-known method to avoid this problem is to use Schwinger's split-point regularization
\begin{align}
  \bar s(0) \gamma_\mu\gamma_5 s(0) &\mapsto  \bar s(x)[x,-x] \gamma_\mu\gamma_5 s(-x)\,,  
\end{align} 
where $x^\mu$ should be sent to zero at the end of the calculation. 
In this case the EOM terms in the divergence can be dropped, but an extra 
contribution appears due to the Wilson line:
\begin{align}
 \partial_\mu \bar s(x) \gamma^\mu\gamma_5 s(-x) &= 2m_s\bar s(x) i\gamma_5 s(-x)
\notag\\& \hspace*{-1cm}
- 2 i \bar s(x) x^\alpha gG_{\alpha\mu}(0)\gamma^\mu\gamma_5 s(-x)\,, 
\end{align} 
cf. Eq.~(\ref{eq:oprel2}). Using the standard expression for the short-distance expansion 
of the quark propagator in a background field~\cite{Novikov:1983gd}
\begin{align}
 \wick{1}{<1 s(-x) >1 {\overline{s}}(x)} &= - \frac{i\slashed{x}}{16\pi^2 x^4} +
 \frac{ix^\rho g\widetilde G_{\rho\sigma}}{16\pi^2 x^2}\gamma^\sigma\gamma_5 +\ldots  
\end{align}
and the symmetric limit $x^\mu\to 0$ such that 
\begin{align}
x_\rho x_\sigma \longrightarrow \frac14 g_{\rho\sigma} x^2, 
\label{symlimit}
\end{align}
one arrives after a little algebra at the expression in (\ref{anomaly}).

The light-ray operators that enter the definitions of DAs are \emph{defined} 
as generating functions of renormalized local operators so that the same
problem with EOM contributions occurs and can be treated in a similar manner. 
We start with a regularized version of the light-ray operator by shifting it
slightly off the light cone
\begin{align}
 \bar s(z_2 n)[z_2n,z_1n] \gamma_\mu\gamma_5 s(z_1n) \mapsto
   \bar s(x_2)[x_2,x_1] \gamma_\mu\gamma_5 s(x_1)
\end{align} 
where
\begin{align}
   x_1 = z_1 n-x\,, &&  x_2 = z_2 n+x\,, && (x\cdot n) =0\,.  
\end{align}
and
\begin{align}
 \Delta^2 &= (x_1-x_2)^2 = x^2\,.   
\end{align}
Then
\begin{eqnarray}
\lefteqn{
\partial_\mu \{\bar q(x_2)\gamma^\mu[x_2, x_1]\gamma_5 q(x_1)\} =}
\nonumber\\
& = & {} + i\int_{0}^1 dv\, \bar q(x_2) \Delta^\alpha
gG_{\alpha\mu}(\bar v x_1 + v x_2 ) \gamma^\mu\gamma_5 q(x_1) 
\nonumber\\&&{}
+ 2m_q\bar q(x_2)i\gamma_5 q(x_1)\,
\label{string1}
\end{eqnarray}
The light-cone expansion of the quark propagator reads~\cite{Balitsky:1987bk} 
\begin{eqnarray}
 \wick{1}{<1 q(x_1) >1 {\overline{q}}(x_2)} &=& \frac{i\slashed{\Delta}}{2\pi^2\Delta^4}[x_1,x_2]
 -
 \frac{\Delta^\rho\gamma^\sigma}{8\pi^2\Delta^2 } \int_0^1du 
\nonumber\\&&{}
\hspace*{-0.2cm}
\times
\Big\{i g\widetilde G_{\rho\sigma}\gamma_5 
+\bar\alpha\alpha (\Delta D)gG_{\rho\sigma}\Big\}(ux_1+\bar u x_2)
\nonumber\\&& +\ldots 
\label{propa}
\end{eqnarray}
where the terms shown by ellipses have at most a logarithmic singularity $\ln\Delta^2 = \ln x^2$ 
and do not contribute in the limit $x\to 0$.

The propagator (\ref{propa}) is traced in (\ref{string1}) with $\gamma^\mu \gamma_5$,
so that only the term in $ig\widetilde G_{\rho\sigma}\gamma_5$ is relevant.
It has a $1/x^2$ singularity, hence we need to collect all contributions with two 
powers of $x$ in the numerator. They can come either from factors of $\Delta$,
that give rise, in the symmetric limit (\ref{symlimit}), to the term
$$
\frac{\alpha_s}{4\pi} \int_0^1\! dv\!\int_0^1\! du\, G^A_{\alpha\mu}(z_{21}^v n) \widetilde G^A_{\alpha\mu}(z_{12}^u n)
$$
or from the expansion of the gluon fields in powers of the deviation from the light-cone direction,
producing contributions of the type
$$
\frac{\alpha_s}{8\pi} z_{12} \int_0^1\! dv\!\int_0^1\! du\,(2v-1)D^\alpha G _{\alpha\mu}(z_{21}^v n) \widetilde G^A_{n\mu}(z_{12}^u n)\,. 
$$
Using the EOM $D^\alpha G^A_{\alpha\mu} = - g\sum_q \bar q t^A \gamma_\mu q$  these contributions can be rewritten in terms of 
the same quark-antiquark-gluon operators that enter Eqs.~(\ref{eq:oprel1}), (\ref{eq:oprel2}), i.e. 
they are of the same order as the NLO $\mathcal{O}(\alpha_s)$ corrections to the coefficient functions of 
twist-four operators. Hence they can (should) be neglected if the calculation is done to LO accuracy.
We obtain
\begin{eqnarray}
\lefteqn{
\partial_\mu \{\bar q(z_1n)\gamma^\mu[z_1n, z_2n]\gamma_5 q(z_2n)\} =}
\nonumber\\
& = & {} - i z_{12} \int_{0}^1 dv\, \bar q(z_1n) n^\alpha
gG_{\alpha\mu}(z_{21}^v n) \gamma^\mu\gamma_5 q(z_2n) 
\nonumber\\&&{}
+ 2m_q\bar q(z_1n)i\gamma_5 q(z_2n)\,
\nonumber\\&&{}
+\frac{\alpha_s}{4\pi} \int_0^1 dv\!\int_0^1 du\, G^A_{\alpha\mu}(z_{21}^v n) \widetilde G^{A;\alpha\mu}(z_{12}^u n)
\end{eqnarray} 
Taking the matrix element of this relation one obtains an equation for the DA $\psi^{(s)}_{4M}(u)$ which 
can be solved as in~\cite{Braun:1989iv,Ball:1998je}
\begin{eqnarray}
 f_{M}^{(s)} \psi_{4M}^{(s)}(u) &=& 2 \phi^{(s)p}_{3M}(u) - 2m^2_M f^{(s)}_M\phi^{(s)}_M(u)
% +  f_{M}^{(s)} \frac{d}{du}
\nonumber\\&&{}\hspace*{-1cm}
+  f_{M}^{(s)} \frac{d}{du}
\int_0^u\! d\alpha_1 \!\int_0^{\bar u}\!d\alpha_2 
\frac{2[\Phi_{4M}(\underline{\alpha})- 2\Psi_{4M}(\underline{\alpha})]}{1-\alpha_1-\alpha_2}
\nonumber\\&&{}\hspace*{-1cm}
+  2 a_M  \delta \psi_{4M}^{(s)}(u)\,,
\label{string2}
\end{eqnarray}  
where the last term $ \delta \psi_{4M}^{(s)}(u)$ is new --- it stems from the anomalous 
contribution in Eq.~(\ref{string2}); $a_M$ is defined in Eq.~(\ref{eq:aM}).

\begin{figure}[t]
\begin{center}
\includegraphics[width=.43\textwidth,clip=true]{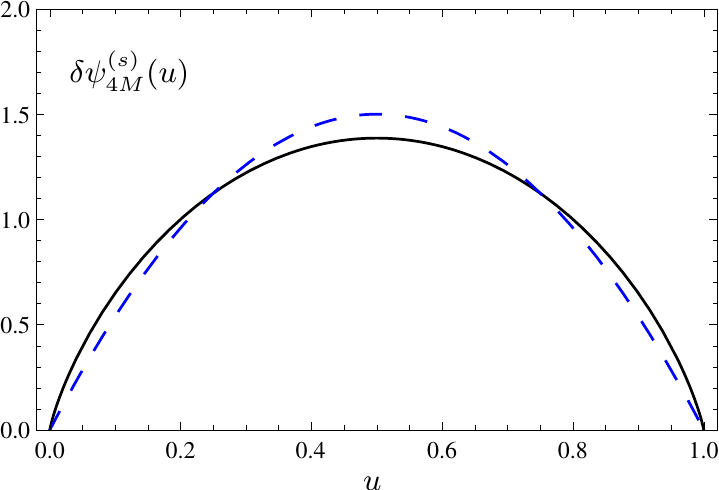}
\end{center}
\caption{The anomalous contribution to the twist-four DA 
$\psi_{4M}^{(s)}(u)$ (\ref{deltapsi}) compared to the 
asymptotic leading-twist DA $6u(1-u)$ (dashed)
}
\label{fig:DAanomaly}
\end{figure}

This extra term can be expressed in terms of the twist-four gluon DA
\begin{align}
\langle 0|\frac{\alpha_s}{4\pi}G(z_2 n)\widetilde{G}(z_1 n) |M(p)\rangle
= a_M \int_0^1 du \, e^{- i z_{21}^u  pn }\phi^{(g)}_{4M}(u)\,, 
%&& \int du\,\phi^{(g)}_{4M}(u)  =1\,, 
%\qquad \qquad \phi_{a}(u) = 6u(1-u)+\ldots
\end{align} 
normalized as $\int du\,\phi^{(g)}_{4M}(u)  =1$.
After some simple algebra one obtains the following equation
for the moments of $\delta \psi_{4M}^{(s)}(u)$:
\begin{eqnarray}
 \lefteqn{\int_0^1 du \,(2u-1)^n\delta \psi_{4M}^{(s)}(u) =}
\\&=&
\frac14 \frac{1+(-1)^n}{(n+1)(n+2)}
 \int_0^1 du\, \Big[ 1- (2u-1)^{n+2}\Big]\frac{\phi^{(g)}_{4M}(u)}{u\bar u}\,,
\nonumber
\end{eqnarray} 
which can be solved for any given twist-four gluon DA. 
A remarkable feature of this equation is that the resulting distribution 
$\delta \psi_{4M}^{(s)}(u)$ depends on the shape of $\phi^{(g)}_{4M}(u)$ only very weakly. Using the asymptotic 
DA $\phi^{(g)}_{4M}(u)=1$ one obtains
\begin{align}
 \delta \psi_{4M}^{(s)}(u) = -2 \Big[ u \ln u + \bar u \ln \bar u\Big],
\label{deltapsi}
\end{align}
whereas for $\phi^{(g)}_{4M}(u)=6u(1-u)$ one gets $\delta \psi_{4M}^{(s)}(u)=6u(1-u)$
as well. The numerical difference between the two expressions is very small, see
Fig.~\ref{fig:DAanomaly}.
 The effect of the anomalous contribution is therefore 
mainly to redefine the normalization of the meson mass correction 
proportional to the twist-two DA, the second term in (\ref{string2}), to
\begin{align}
\lefteqn{
    - 2m^2_M f^{(s)}_M\phi^{(s)}_M(u) +  2 a_M  \delta \psi_{4M}^{(s)}(u)\simeq }
\nonumber\\&\simeq& 
-2(m^2_M f^{(s)}_M -a_M)\phi^{(s)}_M(u) = -2 h^{(s)}_M \phi^{(s)}_M(u)    
\end{align} 
so that it matches the normalization of the pseudoscalar twist-three DA $\phi^{(s)p}_{3M}(u)$
(\ref{TW3:ps}). In this way the condition $\int du\,\psi_{4M}^{(s)}(u) =0$ is restored. 

The complete calculation of such contributions to the twist-four DA is complicated
as it requires reevaluation of all operator identities. Hence relations 
between the parameters, e.g. Eqs.~(\ref{NLOspin}) will be modified. 
This is a large calculation that is beyond the scope of this work.  
Instead, we will assume that the same substitution,
\begin{align}
    m^2_M f^{(s)}_M ~\mapsto~ h^{(s)}_M  = m^2_M f^{(s)}_M -a_M 
\label{ansatz}
\end{align}
can be applied for all occurrences of pseudoscalar meson masses $m^2_M$ in the 
flavor-octet higher-twist corrections.
The ansatz (\ref{ansatz})) is attractive as it guarantees that the higher-twist effects and therefore
also the transition FFs at low momentum transfer obey the same FKS mixing scheme
as is assumed for the leading twist. As we demonstrate in the text, this assumption
does not contradict the existing data.

%%%%%%%%%%%%%%%%%%%%%%%%%%%%%%%%%%%%%%%%%%%%%%%%%%%%%%%%%%%%%%%%%%%%%%%%%%%%%%%%%%%%%%%%%%
%
\section{Scale dependence of the leading-twist DAs to NLO accuracy}
\label{App:RG} 
%
%%%%%%%%%%%%%%%%%%%%%%%%%%%%%%%%%%%%%%%%%%%%%%%%%%%%%%%%%%%%%%%%%%%%%%%%%%%%%%%%%%%%%%%%%%

%%%%%%%%%%%%%%%%%%%%%%%%%%%%%%%%%%%%%%%%%%%%%%%%%%%%%%%%%%%%%%%%%%%%%%%%%%%%%%%%%%%%%%%%%%
%
\subsection{Flavor-octet DAs}
\label{App:octet} 
%
%%%%%%%%%%%%%%%%%%%%%%%%%%%%%%%%%%%%%%%%%%%%%%%%%%%%%%%%%%%%%%%%%%%%%%%%%%%%%%%%%%%%%%%%%%

The scale dependence of the Gegenbauer coefficients in the expansion of the flavor-octet contributions 
to the $\eta,\eta'$ DAs is the same as for the pion DA. One obtains
\cite{Dittes:1983dy,Sarmadi:1982yg,Katz:1984gf,Mikhailov:1984ii,Mueller:1993hg,Mueller:1994cn,Melic:2002ij}
\begin{eqnarray}
 c^{(8)}_n(\mu) &=& c^{(8)}_n(\mu_0)\,  E^{\rm NLO}_n(\mu,\mu_0)
\nonumber\\
&&\hspace*{-1.2cm}{}+
 \frac{\alpha_s(\mu)}{2\pi}\sum_{k=0}^{n-2} c^{(8)}_k(\mu_0)\,  E^{\rm LO}_k(\mu,\mu_0)\,
  d_{n}^{k}(\mu,\mu_0)\,.
\label{eq:NLOevolution-octet}
\end{eqnarray}
The RG factor $E_{n}^{\mathrm{NLO}}(\mu ,\mu _{0})$
in this expression is given by
\begin{eqnarray}
&&E_{n}^{\mathrm{NLO}}(\mu,\mu _{0})=\left[ \frac{\alpha_s(\mu )}{\alpha _{\mathrm{s}}(\mu _{0})}\right] ^{\gamma _{n}^{(0)}/\beta_{0}} 
\\
&&{}\times \left\{ 1 +\frac{\alpha_s(\mu )-\alpha_s(\mu_{0})}{2\pi \beta_0 }
\left(\gamma_{n}^{(1)}-\frac{\beta_{1}}{2\beta_{0}}\gamma _{n}^{(0)}\right) \right\}.
\nonumber
\end{eqnarray}
The corresponding LO RG factor  $E_{n}^{\mathrm{LO}}(\mu ,\mu_{0})$ is obtained by keeping the first term only in the braces.

Here $\beta_0\, (\beta_1)$ and $\gamma_n^{(0)}(\gamma_n^{(1)})$ are the LO (NLO)
coefficients of the QCD $\beta$-function and the anomalous dimensions,
respectively:
\begin{align}
\beta(\alpha_s) = 
\mu^{2}\frac{d\alpha_s}{d\mu^{2}}=
-\alpha_{s} \biggl[ \beta_{0}\frac{\alpha_{s}}{4\pi }
+\beta_{1}\left(\frac{\alpha_s}{4\pi}\right)^2+\ldots\biggr],
\end{align}
\begin{align}
\left[\mu^2 \frac{\partial}{\partial \mu^2} + \beta(\alpha_s)\frac{\partial}{\partial\alpha_s} + \frac12\gamma_n(\alpha_s)\right]
c^{(8)}_n =0\,, 
\notag\\
\gamma_{n}(\alpha _{s}) = \gamma_{n}^{(0)}\frac{\alpha_s}{2\pi }
+\gamma _{n}^{(1)}\left( \frac{\alpha_s}{2\pi}\right)^2+\ldots.
\end{align}
The first two coefficients of the beta-function are
\begin{equation}
\beta _{0}=11-\frac{2}{3}n_{f}\,,\qquad
\beta_{1}=102-\frac{38}{3}n_{f}\,,
\end{equation}%
whereas the LO flavor-nonsinglet anomalous dimensions are given by
\begin{equation}
\gamma _{n}^{(0)} = C_{F}\Big[ 4 \psi(n+2) +4\gamma_E -3 -\frac{2}{(n+1)(n+2)}\Big],
\label{eq:anomdim0}
\end{equation}
where $\psi(x) = d\ln\Gamma(x)/dx$.

The NLO anomalous dimensions can most easily be obtained using the
FeynCalc Mathematica package \cite{FeynCalc}. For convenience we
present explicit expressions for $n=2,4$ that are used in
our calculations ($\gamma_{0}^{(1)}=0$):
\begin{eqnarray}
 \gamma_{2}^{(1)}&=&\frac{17225}{486}-\frac{415}{162}n_{f}\,,
\nonumber\\
 \gamma_{4}^{(1)}&=&\frac{331423}{6750}-\frac{7783}{2025}n_{f}\,. 
\end{eqnarray}
The off-diagonal mixing coefficients $d_n^k$ in Eq.~(\ref{eq:NLOevolution-octet})
are given by the following expression:
\begin{eqnarray}
d_{n}^{k}(\mu ,\mu _{0})&=& r_{n k}(\mu,\mu_{0})\,{M_{n}^{k}}\,, 
\\
r_{n k}(\mu,\mu_{0}) &=& 
\frac{-1}{\gamma_{n}^{(0)}\!-\!\gamma_{k}^{(0)}\!-\!\beta_{0}} 
\biggl\{ 1 \! - \! \left[ \frac{\alpha_{s}(\mu)}{\alpha_{s}(\mu_{0})}\right]^{\frac{\gamma_{n}^{(0)}-\gamma_{k}^{(0)}-\beta_{0}}{\beta_{0}}} \biggr\}. 
\nonumber
\end{eqnarray}
The matrix $M_{n}^{k}$ is defined as
\begin{eqnarray}
M_{n}^{k}&=& \frac{(k+1)(k+2)(2n+3)}{(n+1)(n+2)}\left[\gamma_{k}^{(0)}-\gamma _{n}^{(0)}\right]
\nonumber\\&&\hspace*{-1cm}
\times\left\{ \frac{4C_{F}A_{n}^{k}-\gamma _{k}^{(0)}-\beta _{0}}{(n-k)(n+k+3)}
+2C_F\frac{A_{n}^{k}-\psi (n+2)+\psi (1)}{(k+1)(k+2)}\right\}
\nonumber\\
\label{eq:a.3}
\end{eqnarray}
where
\begin{eqnarray}
A_{n}^{k} &=&\psi \Big( \frac{n+k+4}{2}\Big) -\psi \Big( \frac{n-k}{2}\Big)
\nonumber \\
&&{}+2\psi(n-k)-\psi(n+2)-\psi(1)\,.
\label{eq:a.4}
\end{eqnarray}
%and $\psi(x)= d\ln \Gamma(x)/dx$.
{}For convenience, we give the numerical values of the
nonvanishing coefficients $M_{n}^{k}$ for $n\le 4$:
\begin{eqnarray}
  M_2^0 &=& \frac{455}{162} - \frac{35}{81}n_{f}\,,
\nonumber\\
  M_4^0 &=& \frac{143}{405}- \frac{286}{2025}n_{f}\,,
\nonumber\\
  M_4^2 &=& \frac{6688}{1215} -\frac{836}{2025}n_{f}\,.
\end{eqnarray}

%%%%%%%%%%%%%%%%%%%%%%%%%%%%%%%%%%%%%%%%%%%%%%%%%%%%%%%%%%%%%%%%%%%%%%%%%%%%%%%%%%%%%%%%%%
%
\subsection{Flavor-singlet DAs}
\label{App:singlet} 
%
%%%%%%%%%%%%%%%%%%%%%%%%%%%%%%%%%%%%%%%%%%%%%%%%%%%%%%%%%%%%%%%%%%%%%%%%%%%%%%%%%%%%%%%%%%

The renormalization-group equations for the flavor-singlet quark and gluon DAs 
can be inferred from~\cite{Belitsky:1998uk}. They are more compact 
in matrix notation. To this end we introduce the vector of Gegenbauer coefficients
\begin{eqnarray}
   \vec{c}_n = 
%f^{(1)}_{\eta,\eta'}
\begin{pmatrix} c_n^{(1)}\\ c_n^{(g)}  \end{pmatrix}.
\end{eqnarray} 
Then
\begin{eqnarray}
 \vec{c}_n(\mu) &=& \bit{T}^{-1}_n\bit{E}^{\rm NLO}_n(\mu,\mu_0) \bit{T}_n\vec{c}_n(\mu_0)\,
\\
&&\hspace*{-1.2cm}{}+
 \frac{\alpha_s(\mu)}{2\pi}\sum\limits_{k=0,2,\ldots}^{n-2}\!\!\! \bit{T}^{-1}_n \bit{D}_{n}^{k}(\mu,\mu_0) \bit{E}^{\rm LO}_k(\mu,\mu_0)\bit{T}_k\,
\vec{c}_k(\mu_0)\,\,,
\label{eq:NLOevolution-singlet}
\nonumber
\end{eqnarray}
where $\bit{E}^{\rm NLO(LO)}_n(\mu,\mu_0)$ and $\bit{D}_{n}^{k}(\mu,\mu_0)$ are $2\times 2$ matrices that we will specify 
in what follows and 
\begin{align}
 \bit{T}_n = \text{diag}\left(\frac{3(n+1)(n+2)}{2(2n+3)}, \frac{5n(n+1)(n+2)(n+3)}{24(2n+3)}\right)
\end{align}
is the transformation matrix from the local operator basis of Ref.~\cite{Belitsky:1998uk}
 to the basis of Gegenbauer coefficients
defined in Eqs.~(\ref{phiq}), (\ref{phig}).

Let
\begin{eqnarray}
\boldsymbol{\gamma }_{n}^{(i)}&=&
\begin{pmatrix}
{}^{qq}\gamma _{n}^{(i)} & {}^{qg}\gamma _{n}^{(i)} \\
{}^{gq}\gamma _{n}^{(i)} & {}^{gg}\gamma _{n}^{(i)}
\end{pmatrix}
\end{eqnarray}
be the matrix of anomalous dimensions where the superscript refers to the order 
of perturbation theory. The leading-order expressions are $(n\ge 2$)
\begin{align}
{}^{qq}\gamma _{n}^{(0)} =& C_{F}\Big[ 4 \psi(n+2) +4\gamma_E -3 -\frac{2}{(n+1)(n+2)}\Big],
\notag\\
{}^{qg}\gamma _{n}^{(0)} =& - n_f \frac{12}{(n+1)(n+2)},
\notag\\
{}^{gq}\gamma _{n}^{(0)} =& - C_F \frac{n(n+3) }{3(n+1)(n+2)},
\notag\\
{}^{gg}\gamma _{n}^{(0)} =&  N_c\Big[ 4 \psi(n+2) +4\gamma_E -\frac{8}{(n+1)(n+2)}\Big] -\beta_0.
\end{align} 
The eigenvalues of the LO anomalous dimension matrix $\boldsymbol{\gamma }_{n}^{(0)}$ read:
\begin{eqnarray}
\gamma _{n}^{\pm } &=&\frac{1}{2}\Big[ {}^{qq}\gamma_{n}^{(0)}+{}^{gg}\gamma _{n}^{(0)}
\nonumber\\
&&{}\pm \sqrt{\big( {}^{qq}\gamma _{n}^{(0)}-{}^{gg}\gamma _{n}^{(0)}\big)^{2}+4\, {}^{qg}\gamma _{n}^{(0)}{}^{gq}\gamma _{n}^{(0)}}
\Big].\hspace*{0.7cm}{}
\end{eqnarray}
Then
\begin{align}
\bit{E}^{\rm LO}_n(\mu,\mu_0)
=& 
 \mathbf{P}_{n}^{+} \left[ \frac{\alpha_s(\mu)}{\alpha_s(\mu _{0})}\right]^{\frac{\gamma _{n}^{+}}{\beta_{0}}}
+\mathbf{P}_{n}^{-} \left[  \frac{\alpha _s(\mu)}{\alpha_s(\mu _{0})}\right] ^{\frac{\gamma_{n}^{-}}{\beta_{0}}},
\end{align}
where $\mathbf{P}_n^\pm$ are projectors on the eigenstates of the evolution equation
\begin{align}
& \mathbf{P}_{0}^{+} \,=\,\begin{pmatrix} 1 & 0 \\ 0 & 0 \end{pmatrix},
\qquad \mathbf{P}_{0}^{-} \,=\,\begin{pmatrix} 0 & 0 \\ 0 & 1 \end{pmatrix}, 
\nonumber\\
&\mathbf{P}_{n}^{\pm} \,=\, \pm \frac{1}{\gamma_{n}^{+}-\gamma_{n}^{-}}
\left(\boldsymbol{\gamma}_{n}^{(0)}-\gamma _{n}^{\mp }{\mathbbm 1}\right) ,\quad n\ge 2.
\nonumber\\
&\mathbf{P}_{n}^{+}+\mathbf{P}_{n}^{-}=\mathbf{1},\quad 
(\mathbf{P}_{n}^{\pm})^{2}=\mathbf{P}_{n}^{\pm },\quad
 \mathbf{P}_{n}^{+}\mathbf{P}_{n}^{-}=0.
\end{align}
Further
\begin{align}
\bit{E}_{n}^{\rm NLO}(\mu,\mu_0)
=& \sum\limits_{a,b=\pm} 
\biggl[
\delta_{ab}\mathbf{P}^{a}_{n} +\! \frac{\alpha_s(\mu)}{2\pi} \mathcal{R}_{nn}^{ab} (\mu,\mu_0) \mathbf{P}^{a}_{n}\boldsymbol{\Gamma}_{\!n}\mathbf{P}^{b}_{n}
\biggr]
\notag\\
& \times\, \left[ \frac{\alpha_s(\mu)}{\alpha_s(\mu _{0})}\right]^{\frac{\gamma _{n}^{b}}{\beta_{0}}}
\label{Evol:NLO:Op}
\end{align}
and
\begin{align}
&\bit{D}_{n}^{k}(\mu,\mu_{0}) = \sum_{a,b=\pm} \mathcal{R}_{nk}^{ab}(\mu,\mu_0)\mathbf{P}^{a}_{n}\bit{M}^k_{n}\mathbf{P}^{b}_{k} \,, 
\label{Evol:NLO:OpD}
\end{align}
where 
\begin{align}
\boldsymbol{\Gamma}_{\!n} & =
\boldsymbol{\gamma}_{n}^{(1)}-\frac{\beta_{1}}{2\beta_{0}}\boldsymbol{\gamma}_{n}^{(0)}
\label{Gmatrix}
\end{align}
and
\begin{align}
\mathcal{R}_{n k}^{ab}(\mu,\mu_{0}) = 
\frac{-1}{\gamma_{n}^{a}-\gamma_{k}^{b}-\beta_{0}} 
\biggl\{ 1 \! - \! \left[ \frac{\alpha_{s}(\mu)}{\alpha_{s}(\mu_{0})}\right]^{\frac{\gamma_{n}^{a}-\gamma_{k}^{b}-\beta_{0}}{\beta_{0}}} \biggr\}. 
\label{R:Evol:Op}
\end{align}
The NLO anomalous dimensions matrices for $n=2,4$ are given by~\cite{Mertig:1995ny}
\begin{align}
\boldsymbol{\gamma}_{2}^{(1)} =& 
\begin{pmatrix}
\frac{17225}{486}-\frac{745}{324} n_f {-4 n_f} &  -\frac{43}{216} n_f 
\\[2mm]
 -\frac{7295}{2916}-\frac{25}{243} n_f &  \frac{447}{8}-\frac{437}{81} n_f {-4 n_f} 
\end{pmatrix}, 
\notag\\[2mm]
\boldsymbol{\gamma}_{4}^{(1)} =& 
\begin{pmatrix}
\frac{331423}{6750}-\frac{37963}{10125} n_f {-4 n_f} & \frac{22127}{13500} n_f 
\\[2mm]
-\frac{288421}{91125}-\frac{1316}{6075} n_f & \frac{31744}{375}-\frac{93788}{10125} n_f {-4 n_f} 
\end{pmatrix} 
\end{align}
where the terms $-4n_f$ on the diagonal are due to the factorization of the scale-dependent coupling $f^{(1)}_M$ 
in the definition of the DAs, cf. Eq.~(\ref{eq:anomaly}).  
The matrices $\bit{M}^k_{n}$, $k < n \le 4 $ that describe mixing between different orders in the conformal
(Gegenbauer) expansion are given by 
\begin{align}
\bit{M}^0_2 & =
\begin{pmatrix}
\frac{65}{9}-\frac{4}{9}n_f  & 32 n_f - 6 \pi^2 n_f \\[2mm]
-\frac{175}{27}-\frac{10}{27}n_f & - 1080 + 120 \pi^2 -\frac{10}{3}n_f
\end{pmatrix},
\nonumber \\
\bit{M}^0_4 & =
\begin{pmatrix}
\frac{13}{9}-\frac{14}{45}n_f       & \frac{226}{5}n_f - 6\pi^2 n_f \\[2mm]
-\frac{1414}{135}-\frac{56}{135}n_f & 399 \pi^2 - \frac{18753}{5}-\frac{56}{15}n_f
\end{pmatrix},
\nonumber \\ 
\bit{M}^2_4& =
\begin{pmatrix}
\frac{2128}{243}-\frac{259}{405}n_f    & \frac{49}{30}n_f \\[2mm]
-\frac{4214}{1215}-\frac{196}{1215}n_f & \frac{539}{15}-\frac{98}{405}n_f
\end{pmatrix}.
\end{align}

%%%%%%%%%%%%%%%%%%%%%%%%%%%%%%%%%%%%%%%%%%%%%%%%%%%%%%%%%%%%%%%%%%%%%%%%%%%%%%%%%%%%%%

%%%%%%%%%%%%%%%%%%%%%%%%%%%%%%%%%%%%%%%%%%%%%%%%%%%%%%%%%%%%%%%%%%%%%%%%%%%%%%%%%%%%%%%%%%

%%%%%%%%%%%%%%%%%%%%%%%%%%%%%%%%%%%%%%%%%%%%%%%%%%%%%%%%%%%%%%%%%%%%%%%%%%%%%%%%%%%%%%%%%%


\begin{thebibliography}{99}
\bibitem{footnote1}
Strictly speaking contributions of heavy quarks have to be added at the corresponding thresholds so that 
$n_f \alpha_s(\mu_0) \mapsto 3 \alpha_s(\mu_0) + \alpha_s(\mu_c) + \alpha_s(\mu_b) +\ldots$. Numerically the
difference is not significant.
\end{thebibliography}
\end{document}